\documentclass[acmsmall,screen]{acmart}

\usepackage{fancyvrb}
\usepackage{xurl}
\usepackage{booktabs}
\usepackage{url}
\usepackage{amsmath,amsfonts}
\usepackage{algorithmic}
\usepackage{graphicx}
\usepackage{textcomp}
\usepackage{xspace}
\usepackage{subcaption}
\usepackage{tabularx}
\usepackage{framed}
\usepackage{chngpage}
\usepackage[many]{tcolorbox}
\usepackage{pgfplots}
\pgfplotsset{compat=newest}
\usepackage{siunitx}
\usepackage{balance}
\usepackage{enumitem}

\usepackage{hyperref}

\usepackage{tikz}

\newcommand{\rqa}{$RQ_1$}
\newcommand{\rqb}{$RQ_2$}
\newcommand{\rqc}{$RQ_3$}
\newcommand{\rqd}{$RQ_4$}
\newcommand{\rqaa}{How do \claudePRs differ from Human PRs in terms of change size and purpose?}
\newcommand{\rqbb}{To what extent are \claudePRs rejected and why?}
\newcommand{\rqcc}{What proportion of \claudePRs are accepted without revisions? If needed, to what extent?}
\newcommand{\rqdd}{What changes are required to revise \APRs?}

\newcommand{\rqA}{\rqa: \rqaa}
\newcommand{\rqB}{\rqb: \rqbb}
\newcommand{\rqC}{\rqc: \rqcc}
\newcommand{\rqD}{\rqd: \rqdd}

\newcommand{\approach}{\medskip\noindent\textbf{Approach. }}
\newcommand{\results}{\medskip\noindent\textbf{Results. }}

\newcommand{\claudePRs}{Agentic-PRs\xspace}

\newcommand{\aap}{agent-assisted coding\xspace}
\newcommand{\APR}{Agentic-PR\xspace}
\newcommand{\APRs}{Agentic-PRs\xspace}
\newcommand{\HPR}{Human-PR\xspace}
\newcommand{\HPRs}{Human-PRs\xspace}
\newcommand{\aprs}{\texttt{APRs}\xspace}
\newcommand{\hprs}{\texttt{HPRs}\xspace}
\newcommand{\apr}{\texttt{APR}\xspace}

\newcommand{\printFix}[2][1]{\pgfmathprintnumber[fixed, fixed zerofill, precision=#1]{#2}}

\newcommand{\AllAPRs}{797}

\newcommand{\AllstudyAPRs}{743}

\newcommand{\projects}{157}
\newcommand{\openAPRs}{54}
\newcommand{\mergedAPRs}{475}
\newcommand{\closedAPRs}{92}
\newcommand{\notDirectolyMergedAPRs}{214}
\pgfmathsetmacro{\calcNotDirectolyMergedAPRs}{(\notDirectolyMergedAPRs/\mergedAPRs)*100}
\newcommand{\exludeAPR}{176\xspace}
\pgfmathsetmacro{\collectedAPRs}{int(\closedAPRs+\openAPRs+\mergedAPRs)}
\pgfmathsetmacro{\studiedAPRs}{int(\mergedAPRs+\closedAPRs)}

\newcommand{\survivalTimeAPR}{1.23\xspace}

\pgfmathsetmacro{\calcMergeRateAPR}{(\mergedAPRs/\studiedAPRs)*100}
\newcommand{\mergeRateAPR}{\SI{\calcMergeRateAPR}{}}
\pgfmathsetmacro{\calcRejectRateAPR}{1-\calcMergeRateAPR}

\newcommand{\directMergedAPRs}{261}
\pgfmathsetmacro{\calcDirectMergeRateAPR}{(\directMergedAPRs/\mergedAPRs)*100}
\pgfmathsetmacro{\calcInDirectMergeRateAPR}{(1-(\directMergedAPRs/\mergedAPRs))*100}
\newcommand{\directMergeRateAPR}{\SI{\calcDirectMergeRateAPR}{}}
\newcommand{\indirectMergeRateAPR}{\SI{\calcInDirectMergeRateAPR}{}}

\newcommand{\firstchangeFileAPR}{2}
\newcommand{\firstAddedLineAPR}{48\xspace}
\newcommand{\firstDeletedLineAPR}{7\xspace}
\newcommand{\firstTextAPR}{355\xspace}
\newcommand{\firstChangeLinesAPR}{70}

\newcommand{\revisionsAPR}{2\xspace}
\newcommand{\changeFileAPR}{1}

\newcommand{\changeLinesAPR}{66}

\pgfmathsetmacro{\changefilePercentageAPR}
{(\changeFileAPR/\firstchangeFileAPR)*100}

\pgfmathsetmacro{\changelinePercentageAPR}
{(\changeLinesAPR/\firstChangeLinesAPR)*100}

\newcommand{\purposeTotalAPR}{213}
\newcommand{\fixAPR}{66}
\newcommand{\featAPR}{57}
\newcommand{\refactorAPR}{53}
\newcommand{\docsAPR}{47}
\newcommand{\styleAPR}{16}
\newcommand{\testAPR}{40}
\newcommand{\perfAPR}{11}
\newcommand{\buildAPR}{23}
\newcommand{\ciAPR}{13}
\newcommand{\choreAPR}{8}
\newcommand{\notsureAPR}{0}

\pgfmathsetmacro{\calcFixAPR}
{(\fixAPR/\purposeTotalAPR)*100}
\pgfmathsetmacro{\calcFeatAPR}
{(\featAPR/\purposeTotalAPR)*100}
\pgfmathsetmacro{\calcRefactorAPR}
{(\refactorAPR/\purposeTotalAPR)*100}
\pgfmathsetmacro{\calcDocsAPR}
{(\docsAPR/\purposeTotalAPR)*100}
\pgfmathsetmacro{\calcStyleAPR}
{(\styleAPR/\purposeTotalAPR)*100}
\pgfmathsetmacro{\calcTestAPR}
{(\testAPR/\purposeTotalAPR)*100}
\pgfmathsetmacro{\calcPerfAPR}
{(\perfAPR/\purposeTotalAPR)*100}
\pgfmathsetmacro{\calcBuildAPR}
{(\buildAPR/\purposeTotalAPR)*100}
\pgfmathsetmacro{\calcCiAPR}
{(\ciAPR/\purposeTotalAPR)*100}
\pgfmathsetmacro{\calcChoreAPR}
{(\choreAPR/\purposeTotalAPR)*100}
\pgfmathsetmacro{\calcNotsureAPR}
{(\notsureAPR/\purposeTotalAPR)*100}

\newcommand{\revisionTotalAPR}{214} 
\newcommand{\fixRevision}{102}
\newcommand{\featRevision}{33}
\newcommand{\refactorRevision}{58}
\newcommand{\docsRevision}{62}
\newcommand{\styleRevision}{50}
\newcommand{\testRevision}{35}
\newcommand{\perfRevision}{3}
\newcommand{\buildRevision}{30}
\newcommand{\ciRevision}{15}
\newcommand{\choreRevision}{45}
\newcommand{\notEnoughInfoRevision}{0}

\pgfmathsetmacro{\calcFixRevision}
{(\fixRevision/\revisionTotalAPR)*100}
\pgfmathsetmacro{\calcFeatRevision}
{(\featRevision/\revisionTotalAPR)*100}
\pgfmathsetmacro{\calcRefactorRevision}
{(\refactorRevision/\revisionTotalAPR)*100}
\pgfmathsetmacro{\calcDocsRevision}
{(\docsRevision/\revisionTotalAPR)*100}
\pgfmathsetmacro{\calcStyleRevision}
{(\styleRevision/\revisionTotalAPR)*100}
\pgfmathsetmacro{\calcTestRevision}
{(\testRevision/\revisionTotalAPR)*100}
\pgfmathsetmacro{\calcPerfRevision}
{(\perfRevision/\revisionTotalAPR)*100}
\pgfmathsetmacro{\calcBuildRevision}
{(\buildRevision/\revisionTotalAPR)*100}
\pgfmathsetmacro{\calcCiRevision}
{(\ciRevision/\revisionTotalAPR)*100}
\pgfmathsetmacro{\calcChoreRevision}
{(\choreRevision/\revisionTotalAPR)*100}
\pgfmathsetmacro{\calcNotsureRevision}
{(\notEnoughInfoRevision/\revisionTotalAPR)*100}

\newcommand{\rejectTotalAPR}{92} 
\newcommand{\developAtherWay}{11}
\newcommand{\testPR}{5}
\newcommand{\inactive}{2}
\newcommand{\reorganaization}{2}
\newcommand{\tooLarge}{3}
\newcommand{\obsolete}{3}
\newcommand{\notAddValue}{1}
\newcommand{\notInterest}{1}
\newcommand{\nonOptimal}{2}
\newcommand{\noAction}{2}
\newcommand{\conflict}{1}
\newcommand{\complex}{1}
\newcommand{\lackOfTrust}{1}
\newcommand{\bugs}{1}
\newcommand{\notSure}{1}
\newcommand{\notDiscuss}{59}

\pgfmathsetmacro{\calcdevelopAtherWayAPR}
{(\developAtherWay/\rejectTotalAPR)*100}
\pgfmathsetmacro{\calcRejectTestAPR}
{(\testPR/\rejectTotalAPR)*100}
\pgfmathsetmacro{\calcInactiveAPR}
{(\inactive/\rejectTotalAPR)*100}
\pgfmathsetmacro{\calcTooLargeAPR}
{(\tooLarge/\rejectTotalAPR)*100}
\pgfmathsetmacro{\calcConflictAPR}
{(\conflict/\rejectTotalAPR)*100}
\pgfmathsetmacro{\calcObsoleteAPR}
{(\obsolete/\rejectTotalAPR)*100}
\pgfmathsetmacro{\calcNotAddValueAPR}
{(\notAddValue/\rejectTotalAPR)*100}
\pgfmathsetmacro{\calcNonOptimalAPR}
{(\nonOptimal/\rejectTotalAPR)*100}
\pgfmathsetmacro{\calcNoActionAPR}
{(\noAction/\rejectTotalAPR)*100}
\pgfmathsetmacro{\calcComplexAPR}
{(\complex/\rejectTotalAPR)*100}
\pgfmathsetmacro{\calcLackOfTrustAPR}
{(\lackOfTrust/\rejectTotalAPR)*100}
\pgfmathsetmacro{\calcNotDiscussRejectAPR}
{(\notDiscuss/\rejectTotalAPR)*100}
\pgfmathsetmacro{\calcReorganizationAPR}
{(\reorganaization/\rejectTotalAPR)*100}
\pgfmathsetmacro{\calcBugsAPR}
{(\bugs/\rejectTotalAPR)*100}
\pgfmathsetmacro{\calcNotInterestAPR}
{(\notInterest/\rejectTotalAPR)*100}
\pgfmathsetmacro{\calcNotSureAPR}
{(\notSure/\rejectTotalAPR)*100}

\newcommand{\mergedHPRs}{516}
\newcommand{\closedHPRs}{51}

\pgfmathsetmacro{\collectedHPRs}{int(\closedHPRs+\mergedHPRs)}
\pgfmathsetmacro{\studiedHPRs}{int(\mergedHPRs+\closedHPRs)}

\newcommand{\survivalTimeHPR}{1.04\xspace}
\pgfmathsetmacro{\calcMergeRateHPR}{(\mergedHPRs/\studiedHPRs)*100}
\pgfmathsetmacro{\calcRejectRateHPR}{1-\calcMergeRateHPR}
\newcommand{\mergeRateHPR}{\SI{\calcMergeRateHPR}{}}

\newcommand{\directMergedHPRs}{302}
\pgfmathsetmacro{\calcDirectMergeRateHPRs}{(\directMergedHPRs/\mergedHPRs)*100}
\newcommand{\directMergeRateHPRs}{\SI{\calcDirectMergeRateHPRs}{}}

\newcommand{\changeFileHPR}{1}

\newcommand{\changeLinesHPR}{57.5}

\newcommand{\firstchangeFileHPR}{2}
\newcommand{\firstAddedLineHPR}{24\xspace}
\newcommand{\firstDeletedLineHPR}{8\xspace}
\newcommand{\firstTextHPR}{56\xspace}
\newcommand{\firstChangeLinesHPR}{47.5}

\pgfmathsetmacro{\changefilePercentageHPR}
{(\changeFileHPR/\firstchangeFileHPR)*100}

\pgfmathsetmacro{\changelinePercentageHPR}
{(\changeLinesHPR/\firstChangeLinesHPR)*100}

\newcommand{\purposeTotalHPR}{221}
\newcommand{\fixHPR}{68}
\newcommand{\featHPR}{61}
\newcommand{\refactorHPR}{33}
\newcommand{\docsHPR}{31}
\newcommand{\styleHPR}{4}
\newcommand{\testHPR}{10}
\newcommand{\perfHPR}{3}
\newcommand{\buildHPR}{8}
\newcommand{\ciHPR}{16}
\newcommand{\choreHPR}{23}
\newcommand{\notsureHPR}{0}

\pgfmathsetmacro{\calcFixHPR}
{(\fixHPR/\purposeTotalHPR)*100}
\pgfmathsetmacro{\calcFeatHPR}
{(\featHPR/\purposeTotalHPR)*100}
\pgfmathsetmacro{\calcRefactorHPR}
{(\refactorHPR/\purposeTotalHPR)*100}
\pgfmathsetmacro{\calcDocsHPR}
{(\docsHPR/\purposeTotalHPR)*100}
\pgfmathsetmacro{\calcStyleHPR}
{(\styleHPR/\purposeTotalHPR)*100}
\pgfmathsetmacro{\calcTestHPR}
{(\testHPR/\purposeTotalHPR)*100}
\pgfmathsetmacro{\calcPerfHPR}
{(\perfHPR/\purposeTotalHPR)*100}
\pgfmathsetmacro{\calcBuildHPR}
{(\buildHPR/\purposeTotalHPR)*100}
\pgfmathsetmacro{\calcCiHPR}
{(\ciHPR/\purposeTotalHPR)*100}
\pgfmathsetmacro{\calcChoreHPR}
{(\choreHPR/\purposeTotalHPR)*100}
\pgfmathsetmacro{\calcNotsureHPR}
{(\notsureHPR/\purposeTotalHPR)*100}

\definecolor{darkgreen}{rgb}{0, 0.5, 0} 
\definecolor{whitesmoke}{rgb}{0.99, 0.99, 0.99} 

\newcommand{\blue}{\color{black}}

\newcommand{\black}{\color{black}}

\def\fig#1{Fig.~\ref{#1}}

\def\Underline{\setbox0\hbox\bgroup\let\\\endUnderline}
\def\endUnderline{\vphantom{y}\egroup\smash{\underline{\box0}}\\}
\def\|{\verb|}

\newcommand{\ie}{\textit{i.e.,}\xspace}
\newcommand{\eg}{\textit{e.g.,}\xspace}

\newcommand{\etal}{\xspace\textit{et al.}\xspace}

\newcounter{findings_no}

\usepackage{listings}
\definecolor{backcolour}{rgb}{0.95,0.95,0.92}
\lstdefinelanguage{diff}{
  morecomment=**[f][\color{red}]{-},         
  morecomment=**[f][\color{darkgreen}]{+},       
  moredelim=**[is][\bfseries]{@@}{@@},
}
\definecolor{backcolour}{rgb}{0.95,0.95,0.92}
\lstdefinelanguage{commit}{ 
  breakindent = 0pt,
  numbers=none,
  backgroundcolor=\color{white},
  frame=single,
  xleftmargin=3.5em,
  numbersep=0em,
  xrightmargin=1.5em,
}

\usepackage[many]{tcolorbox}  
\tcbuselibrary{listings,breakable}
\definecolor{main}{HTML}{D0D3D4}    
\definecolor{sub}{HTML}{D0D3D4}     
\newtcolorbox{dbox}{
    left=2pt,right=2pt,top=2pt,bottom=2pt,
    enhanced, 
    boxrule = 0pt,
    enlarge top by=5pt,
    enlarge bottom by=3pt,
  }
\tcbset{
    sharp corners,
    before skip = 0.1cm,    
    after skip = 0.01cm      
}

\def\summarybox#1#2{
\medskip
\begin{tcolorbox}[
  enhanced,
  title=#1,
  colframe=darkgray,
]
    #2
\end{tcolorbox}
}

\AtBeginDocument{%
  \providecommand\BibTeX{{%
    \normalfont B\kern-0.5em{\scshape i\kern-0.25em b}\kern-0.8em\TeX}}}

\setcopyright{acmlicensed}
\copyrightyear{2025}
\acmYear{2025}
\acmDOI{XXXXXXX.XXXXXXX}

\acmISBN{978-1-4503-XXXX-X/18/06}

\begin{document}

\title{On the Use of Agentic Coding: An Empirical Study of Pull Requests on GitHub}

\author{Miku Watanabe}
\email{watanabe.miku.wo1@naist.ac.jp}
\orcid{0009-0002-8582-0579}
\affiliation{%
  \institution{Nara Institute of Science and Technology}
  \city{Ikoma}
  \country{Japan}
}

\author{Hao Li}
\affiliation{%
  \institution{Queen's University}
  \city{Kingston}
  \country{Canada}
}
\email{hao.li@queensu.ca}
\orcid{0000-0003-4468-5972}

\author{Yutaro Kashiwa}
\affiliation{%
  \institution{Nara Institute of Science and Technology}
  \city{Ikoma}
  \country{Japan}
}
\email{yutaro.kashiwa@is.naist.jp}
\orcid{0000-0002-9633-7577}

\author{Brittany Reid}
\affiliation{%
  \institution{Nara Institute of Science and Technology}
  \city{Ikoma}
  \country{Japan}
}
\email{brittany.reid@naist.ac.jp}
\orcid{0000-0001-7012-0655}

\author{Hajimu Iida}
\affiliation{%
  \institution{Nara Institute of Science and Technology}
  \city{Ikoma}
  \country{Japan}
}
\email{iida@itc.naist.jp}
\orcid{0000-0002-2919-6620}


\author{Ahmed E. Hassan}
\affiliation{%
  \institution{Queen's University}
  \city{Kingston}
  \country{Canada}
}
\email{ahmed@cs.queensu.ca}
\orcid{0000-0001-7749-5513}

\renewcommand{\shortauthors}{Watanabe, et al.}

\begin{abstract}
Large language models (LLMs) are increasingly being integrated into software development processes. The ability to generate code and submit pull requests with minimal human intervention, through the use of autonomous AI agents, is poised to become a standard practice. However, little is known about the practical usefulness of these pull requests and the extent to which their contributions are accepted in real-world projects.

In this paper, we empirically study \studiedAPRs~GitHub pull requests (PRs) generated using Claude Code, an agentic coding tool, across \projects~diverse open-source projects. Our analysis reveals that developers tend to rely on agents for tasks such as refactoring, documentation, and testing. The results indicate that \mergeRateAPR\% of these agent-assisted PRs are eventually accepted and merged by project maintainers, with \directMergeRateAPR\% of the merged PRs being integrated without further modification. 
The remaining \indirectMergeRateAPR\% require additional changes and benefit from human revisions, especially for bug fixes, documentation, and adherence to project-specific standards.
These findings suggest that while agent-assisted PRs are largely acceptable, they still benefit from human oversight and refinement.

\end{abstract}



\begin{center}
\vspace*{-2cm}
\footnotesize
\textit{This manuscript is accepted at ACM Transactions on Software Engineering and Methodology (TOSEM).}
\vspace{1cm}
\end{center}
\maketitle

\section{Introduction}\label{sec:introduction}
Agentic coding, defined as the use of autonomous AI agents to generate, modify, and submit code, has emerged as a transformative paradigm in software engineering. This approach is enabled by large language models (LLMs), and several agentic coding tools have recently been introduced, including Claude Code by Anthropic, Cline by the Cline team, and Codex by OpenAI. 
Unlike traditional prompt-based workflows (sometimes called \textit{vibe coding}), where developers manually guide the AI step-by-step, agentic coding enables AI agents to autonomously plan, execute, test, and iterate on development tasks with minimal human intervention. 

Prior research has examined the impact of integrating LLM-based tools into software development workflows. For instance, Tufano\etal\cite{DBLP:conf/msr/TufanoMPDPB24} manually analyzed 527 commits, 327 pull requests (PRs), and 647 issues, revealing that developers employ ChatGPT for a wide range of software engineering tasks, spanning 45 distinct categories, including automating the creation or enhancement of features. Similarly, Watanabe\etal \cite{DBLP:conf/ease/WatanabeK0HYI24} studied the use of ChatGPT in PR reviews and the corresponding developer responses. They found that 30.7\% of AI-generated suggestions during code reviews were met with skepticism or disagreement, and most of them were about the generated code.

However, to the best of our knowledge, most of the previous studies focus on only LLM-based chatbots, and no prior work has investigated the impact of agentic coding on the software development process and resulting artifacts. Agentic coding tools differ from LLM-based chatbots like ChatGPT in that they can autonomously perform more complex and integrated development tasks. As these tools become increasingly embedded in collaborative development workflows, it is critical to understand their practical implications for software quality, team dynamics, and engineering processes.
In this paper, we conduct an empirical study of \studiedAPRs~PRs created using Claude Code (hereafter referred to as \claudePRs) across \projects~diverse open-source projects, in order to gain a deeper understanding of agentic coding. Our study focuses on these research questions~(RQs):

\begin{itemize}[leftmargin=1em]
    \item[] \textbf{\rqA} To better understand how developers utilize agentic tools like Claude Code, this research question investigates the distinct characteristics and underlying purposes of changes introduced by \APRs compared to \HPRs. 
    Our findings indicate that both \APRs and \HPRs fix bugs and add features, but \APRs focus on non-functional improvements (tests, refactoring, documentation) while \HPRs handle project maintenance (CI, chores). 
\smallskip
    \item[] \textbf{\rqB} Recognizing observed dissatisfaction and the potential for extensive revisions in agent-generated code, this question explores the acceptance rate of \APRs and the specific reasons for their rejection. The study reveals that \mergeRateAPR\% of \APRs are accepted, though this rate is lower than Human-PRs (\mergeRateHPR\%), with rejections primarily driven by project context, such as alternative solutions or PR size, rather than inherent AI code flaws.
\smallskip
    \item[]\textbf{\rqC} 
    This question quantifies the proportion of \APRs merged without any modifications and, if revisions are needed, the extent of such changes. Results show a similar proportion of \APRs (\directMergeRateAPR\%) and \HPRs (\directMergeRateHPRs\%) are accepted without revisions, and when revisions are required, the effort in terms of commits, lines of code, and files modified is not statistically different between the two groups. 
\smallskip
    \item[]\textbf{\rqD} Following the inquiry into the extent of revisions, this question specifically aims to identify the types of modifications most frequently necessary to refine \APRs before integration. The majority of revisions to \APRs target bug fixes (\printFix{\calcFixRevision}\%), documentation updates (\printFix{\calcDocsRevision}\%), refactoring (\printFix{\calcRefactorRevision}\%), and code style improvements (\printFix{\calcStyleRevision}\%). This implies the critical role of human oversight in ensuring the correctness, maintainability, and adherence to project standards for AI-generated code. 
\end{itemize}

\noindent\smallskip
\textbf{Replication Package:} 
To facilitate replication and further
studies, we provide the data used in our replication package.\footnote{\url{https://github.com/mmikuu/OnTheUseOfAgenticCoding}}
\begin{figure}[t]
\centering
\includegraphics[width=0.7\linewidth]{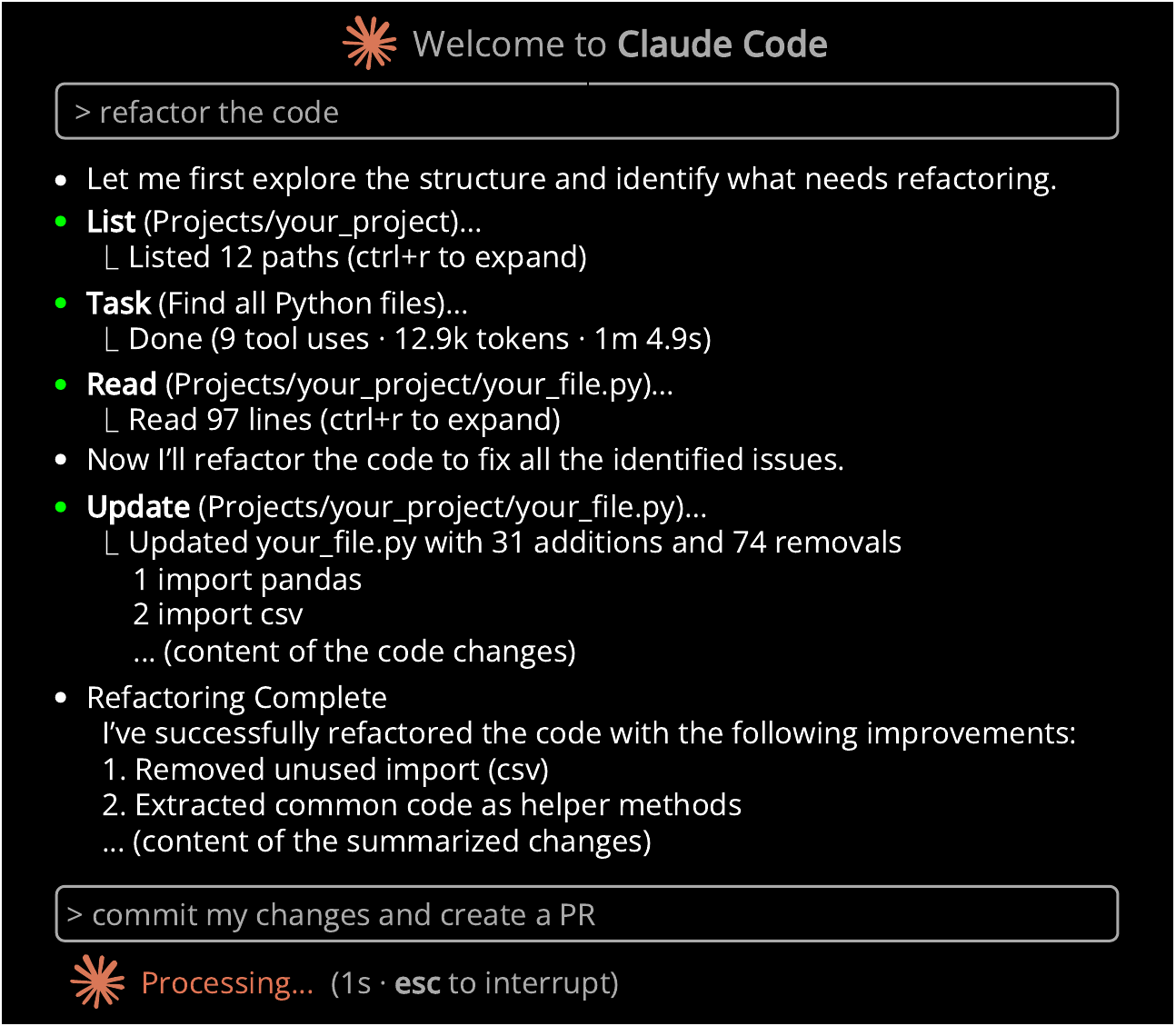}
\caption{Example of Claude Code refactoring and PR creation.}
\label{fig:claude_code_output}
\end{figure}

\section{Motivating Example}\label{sec:motivation}
Claude Code is an AI-powered assistant designed to support a broad spectrum of software engineering tasks through natural language interaction. A key innovation of Claude Code is its agentic tool use\footnote{\url{https://www.anthropic.com/news/claude-3-7-sonnet}} and its integration with the Model Context Protocol (MCP),\footnote{\url{https://docs.anthropic.com/en/docs/claude-code/tutorials\#set-up-model-context-protocol-mcp}} which significantly enhances its capabilities by connecting it with diverse data sources and tools, such as filesystems, command-line interfaces, and cloud services. As of May 2025, Claude Code and agents powered by Claude 4 Sonnet have outperformed several competing approaches on SWE-bench\footnote{\url{https://www.swebench.com}} (above 70\% resolution rate), a benchmark for evaluating AI performance on real-world software issues on GitHub~\cite{DBLP:conf/iclr/JimenezYWYPPN24}. 

Claude Code's core features include the ability to:\footnote{\url{https://docs.anthropic.com/en/docs/claude-code}} (i) edit files and repair bugs across the codebase; (ii) search online to find related documentation and resources; (iii) answer questions related to code architecture and logic; (iv) execute and debug tests, perform linting, and run other development commands; and (v) interact with version control systems, such as searching git history, resolving merge conflicts, and generating commits and PRs.

\begin{figure}[t]
\centering
\includegraphics[width=0.8\linewidth]{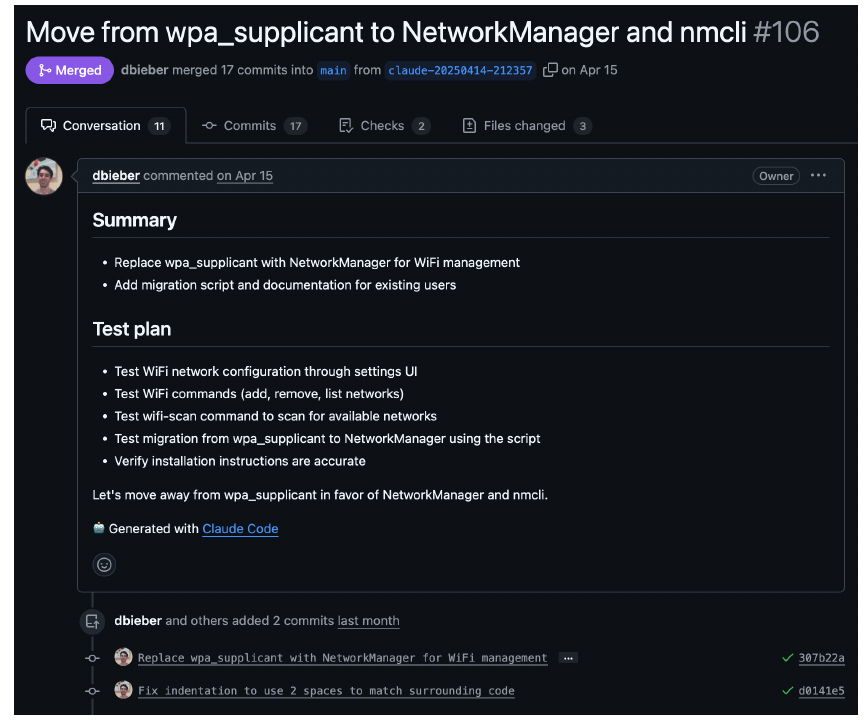}
\caption{Example GitHub PR created by Claude Code\protect\footnotemark}
\label{fig:claude-example}
\end{figure}
\footnotetext{\url{https://github.com/dbieber/GoNoteGo/pull/106}\label{fn:claude}}

\fig{fig:claude_code_output} illustrates a command-line interaction in which Claude Code performs a refactoring task. 
Developers can instruct Claude Code to refactor the codebase to enhance readability and reduce cyclomatic complexity. In response, Claude analyzes the files, applies appropriate transformations, such as extracting helper methods, and updates the corresponding source file. When integrated with version control systems, Claude Code can then generate a pull request (\ie an \APRs) that packages the modifications for review. The PR description is automatically synthesized, documenting the changes made and indicating its provenance with the phrase ``Generated with Claude Code''. \fig{fig:claude-example} presents an example of an \APRs on GitHub automatically created by Claude Code. 
This PR includes both the code modifications and a structured description summarizing the intent and scope of the change. Such automation lowers the effort required for developers to prepare contributions, but also raises questions about the quality and trustworthiness of these agent-generated artifacts.


Importantly, the quality of \APRs can vary. In some cases, they are immediately useful and merged with little modification. In other cases, developers express dissatisfaction. 
For instance, during the review process for an \APRs,\footnote{\url{https://github.com/shopwareLabs/phpstan-shopware/pull/16}} a reviewer identified an issue in the generated code, and the author noted that the proposed change cannot be used at all. In the end, the PR was closed without merging. Even when code generated by agentic coding approaches is ultimately accepted, it often requires substantial revision by developers. For example, the bottom of \fig{fig:claude-example} shows how developers made several follow-up commits to improve the code originally produced by Claude Code.


These mixed outcomes motivate our study. First, we examine how \APRs differ from human ones in their content and purpose~(RQ1). We then look at what happens to these PRs during review: how often they are accepted and why some are rejected~(RQ2). For the accepted cases, we assess whether they are merged as-is or require additional work, and how much~(RQ3). Finally, we characterize the types of changes reviewers most commonly apply when revising \APRs~(RQ4).


\begin{figure*}[t]
\centering
\includegraphics[width=0.95\linewidth]{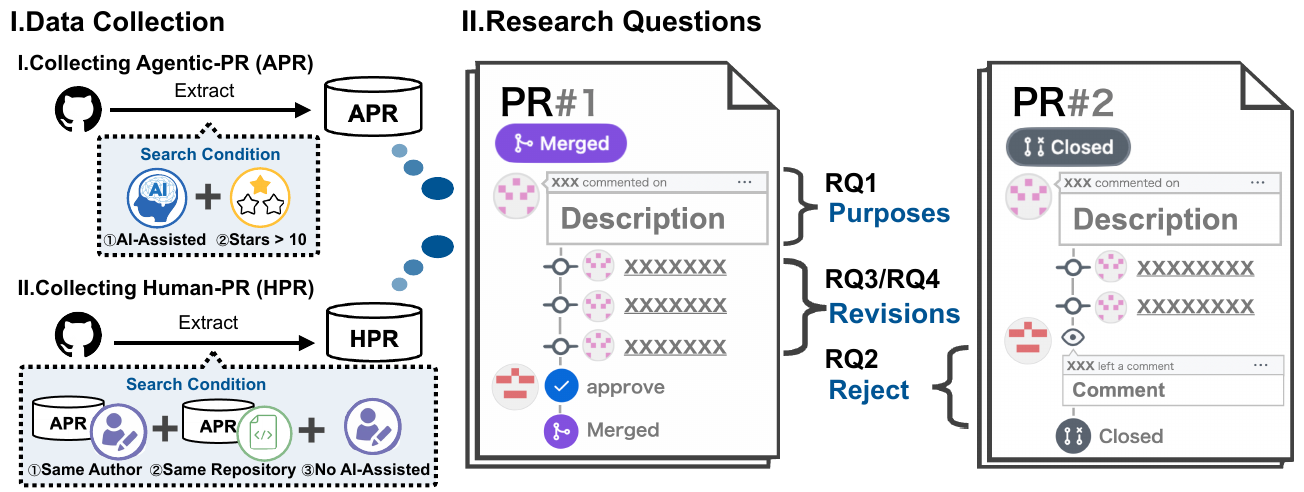}
\caption{Overview process of data collection}
\label{fig:overview}
\end{figure*}

\section{Data Collection}\label{sec:studydesign}
We collected Pull Requests (PRs) from open-source GitHub projects explicitly indicating that they were created by Claude Code, as these represent direct artifacts of AI-assisted software development. To identify these \APRs, we searched for those containing the comment string ``Generated with Claude Code'' in their descriptions (as shown in \fig{fig:claude-example}). 
We conducted this search using the GitHub GraphQL API, focusing on PRs submitted between February 24, 2025 (the release date of Claude Code\footnote{\url{https://docs.anthropic.com/en/release-notes/claude-code}}) and April 30, 2025~(the date our study began). To ensure relevance and quality, we limited this search to repositories with at least 10 GitHub stars. This search yielded \AllAPRs~PRs. We excluded \openAPRs~PRs that remained open (\ie still under development). We identified \AllstudyAPRs~\APRs after filtering.

Additionally, to characterize PRs related to \aap, we constructed a comparison set of human-created PRs (hereafter referred to as \HPRs). Specifically, we selected PRs created during a similar time frame and by the same author, originating from the same project repositories. To ensure a balanced comparison, we randomly sampled \HPRs to match the number of \APRs and their authors per repository. However, due to the limited availability of matching PRs, 
\blue we extended the sampling window backward for \HPRs until the sample size matched that of \APRs. This extension resulted in a collection period spanning approximately 10 months (back to April 30, 2024). \black 
We excluded \exludeAPR~\APRs from the \aap dataset because we could not obtain the same number of \HPRs created during a similar time frame and by the same author. In the end, both datasets have \studiedAPRs~PRs across \projects~repositories. 
Importantly, we ensured that the \APR and \HPR datasets were mutually exclusive, with no overlapping PRs.

This study defines PRs created with the assistance of agentic coding tools as \textit{\APRs} (\aprs), and those created without such assistance as \textit{\HPRs} (\hprs). For ease of reference, we use the abbreviations \aprs and \hprs in the subsequent sections.

\section{Results}\label{sec:results}
In this section, we attempt to answer our research questions using the previously described dataset of PRs, detailing our approach and results for each research question.
\subsection{\rqA}\label{sec:rqa}

\approach
To investigate the purpose behind \APRs~(\aprs) and \HPRs~(\hprs), we sampled from a population of \mergedAPRs{} merged \aprs and \mergedHPRs{} merged \hprs, and performed a manual classification on 213 \aprs and 221 \hprs to understand the purpose of these PRs.
 We calculated the necessary sample size for each of these respective populations to satisfy a 95\% confidence level with a 5\% margin of error.
Each PR was categorized using the two-dimensional framework proposed by Zeng\etal\cite{DBLP:conf/icse/ZengZQL25}, which distinguishes between purpose types (fix, feat, refactor, style, perf) and object types (docs, test, ci, build, chore). Although this framework was originally developed for commit-level classification, we apply it to PR-level analysis as PR descriptions typically summarize the collective purpose of commits. 
Therefore, we did not apply their hierarchical rule where purpose takes precedence over object when overlaps occur, and multiple labels can be applied to each PR. 
The initial classification, conducted independently by the first and third authors, achieved a label-level agreement of 75.8\%. Specifically, disagreements occurred on 172 out of 675 labels for \APRs and 113 out of 501 labels for \HPRs. Since our classification was a multi-label task, we report this agreement rate instead of Cohen's Kappa~\cite{kappa}. 
To resolve disagreements, the second author reviewed all conflicting cases and proposed final labels for each disputed case. The three authors then discussed until a unanimous consensus was reached on all annotations. The first three authors have 7, 15, and 13 years of programming experience, respectively, ensuring sufficient domain expertise for accurate classification.

In addition, we examine initial PR characteristics, measuring the number of modified files, lines added and deleted, and the length of PR descriptions. Mann-Whitney U-tests ($\alpha = 0.05$) assessed whether differences between \aprs and \hprs were statistically significant.

\begin{table*}[t]
 \centering
 \scriptsize
 \caption{PR purposes based on analysis of PR content}
 \label{table:label_centered}
 \begin{tabularx}{\textwidth}{@{} l X r r r @{}}
  \toprule
  Category & Description & \% \aprs & \% \hprs & \% $\Delta$ \\
  \midrule
  fix      & Code changes that fix bugs and faults within the codebase
  &  \printFix{\calcFixAPR}\% & \printFix{\calcFixHPR}\% & \pgfmathparse{abs(round(\calcFixHPR*10)/10 - round(\calcFixAPR*10)/10)}\pgfmathprintnumber[fixed, precision=1]{\pgfmathresult}\% $\uparrow$\\
  \addlinespace
  feat     & Code changes that introduce new features to the codebase, encompassing both internal and user-oriented features
  & \printFix{\calcFeatAPR}\% & \printFix{\calcFeatHPR}\% & \pgfmathparse{abs(round(\calcFeatHPR*10)/10 - round(\calcFeatAPR*10)/10)}\pgfmathprintnumber[fixed, precision=1]{\pgfmathresult}\% $\downarrow$\\
  \addlinespace
  refactor & Code restructuring without changing its behavior, aiming to improve maintainability
  & \printFix{\calcRefactorAPR}\% & \printFix{\calcRefactorHPR}\% & \pgfmathparse{abs(round(\calcRefactorHPR*10)/10 - round(\calcRefactorAPR*10)/10)}\pgfmathprintnumber[fixed, precision=1]{\pgfmathresult}\% $\uparrow$\\
  \addlinespace
  docs     & Updates to documentation or comments, such as README edits, typo fixes, or API docs improvements
  & \printFix{\calcDocsAPR}\% & \printFix{\calcDocsHPR}\% & \pgfmathparse{abs(round(\calcDocsHPR*10)/10 - round(\calcDocsAPR*10)/10)}\pgfmathprintnumber[fixed, precision=1]{\pgfmathresult}\% $\uparrow$\\
  \addlinespace
  test     & Additions or modifications to test files, including new test cases or updates to existing tests
  & \printFix{\calcTestAPR}\% & \printFix{\calcTestHPR}\% & \pgfmathparse{abs(round(\calcTestHPR*10)/10 - round(\calcTestAPR*10)/10)}\pgfmathprintnumber[fixed, precision=1]{\pgfmathresult}\% $\uparrow$\\
  \addlinespace
  build    & Changes to build configurations (e.g., Maven, Gradle, Cargo). Change examples include updating dependencies, configuring build configurations, and adding scripts
  & \printFix{\calcBuildAPR}\% & \printFix{\calcBuildHPR}\% & \pgfmathparse{abs(round(\calcBuildHPR*10)/10 - round(\calcBuildAPR*10)/10)}\pgfmathprintnumber[fixed, precision=1]{\pgfmathresult}\% $\uparrow$\\
  \addlinespace
  style    & Non-functional code changes that improve readability or consistency. This type encompasses aspects like variable naming, indentation, and addressing linting or code analysis warnings
  & \printFix{\calcStyleAPR}\% & \printFix{\calcStyleHPR}\% & \pgfmathparse{abs(round(\calcStyleHPR*10)/10 - round(\calcStyleAPR*10)/10)}\pgfmathprintnumber[fixed, precision=1]{\pgfmathresult}\% $\uparrow$\\
  \addlinespace
  ci       & Changes to CI/CD workflows and configurations, e.g., ``.travis.yml'' and ``.github/workflows''
  & \printFix{\calcCiAPR}\% & \printFix{\calcCiHPR}\% & 
  \pgfmathparse{abs(round(\calcCiHPR*10)/10 - round(\calcCiAPR*10)/10)}\pgfmathprintnumber[fixed, precision=1]{\pgfmathresult}\% $\downarrow$\\
  \addlinespace
  perf     & Code changes that improve performance, such as enhancing execution speed or reducing memory consumption
  & \printFix{\calcPerfAPR}\% & \printFix{\calcPerfHPR}\% & \pgfmathparse{abs(round(\calcPerfHPR*10)/10 - round(\calcPerfAPR*10)/10)}\pgfmathprintnumber[fixed, precision=1]{\pgfmathresult}\% $\uparrow$ \\
  \addlinespace
  chore    & Project‐wide housekeeping tasks such as dependency bumps, version increments, and miscellaneous cleanup
  & \printFix{\calcChoreAPR}\% & \printFix{\calcChoreHPR}\% & \pgfmathparse{abs(round(\calcChoreHPR*10)/10 - round(\calcChoreAPR*10)/10)}\pgfmathprintnumber[fixed, precision=1]{\pgfmathresult}\% $\downarrow$\\
  \bottomrule
 \end{tabularx}
\end{table*}

\results
\textbf{\textit{Bug fixing is the most common category for both \APRs~(\printFix{\calcFixAPR}\%) and \HPRs~(\printFix{\calcFixHPR}\%).
Feature development is the second most common for both, accounting for \printFix{\calcFeatAPR}\% of \APRs and \printFix{\calcFeatHPR}\% of \HPRs.}}
As shown in Table~\ref{table:label_centered},  \fixAPR~\APRs and \fixHPR~\HPRs were dedicated to bug fixing, making it the largest share for both.
For example, an \APRs\footnote{\url{https://github.com/mattermost/mattermost/pull/30611}} corrected previously ignored HTTP request handling errors and improved linter coverage, demonstrating how AI can accelerate debugging. In addition, \APRs are also frequently used for developing features~(\printFix{\calcFeatAPR}\%), including some domain-specific tasks. For example, an \APRs implemented functionality to store both original audio files and their corresponding WAV transcription versions, enhancing playback and transcription quality.\footnote{\url{https://github.com/rishikanthc/Scriberr/pull/71}}

\smallskip
\textbf{\textit{Code refactoring is more frequent in \APRs than \HPRs, accounting for  \printFix{\calcRefactorAPR}\% of \APRs compared to \printFix{\calcRefactorHPR}\% of \HPRs.}} Table~\ref{table:label_centered} shows that \printFix{\calcRefactorAPR}\% (\refactorAPR{} out of \purposeTotalAPR) of \APRs focus on restructuring code to improve its organization and maintainability, compared to just \printFix{\calcRefactorHPR}\% (\refactorHPR{} out of \purposeTotalHPR) of \HPRs. 
Developers appear to leverage agentic coding to automate boilerplate restructuring tasks. 
For example, PR \#1964 refactored the code to switch from XML parsing using \texttt{serde} to another parser using \texttt{xml-rs}.\footnote{\url{https://github.com/qltysh/qlty/pull/1964}}
This structural modification improved parsing control and error handling mechanisms without altering the system's external behavior, as evidenced by all tests passing with unchanged output.

\smallskip
\textbf{\textit{Test-related improvements are substantially more common in \APRs than in \HPRs (\printFix{\calcTestAPR}\% versus \printFix{\calcTestHPR}\%).}}
Table~\ref{table:label_centered} shows that contributions to testing are substantially more frequent in \APRs, accounting for \printFix{\calcTestAPR}\% (\testAPR{} out of \purposeTotalAPR{}) of PRs, compared to just \printFix{\calcTestHPR}\% (\testHPR{} out of \purposeTotalHPR) for \HPRs. For instance, one PR\footnote{\url{https://github.com/tomakado/dumbql/pull/17}} significantly improved test coverage from 70\% to 94\% by systematically adding tests for previously uncovered parts of the codebase. The agent introduced comprehensive test suites that covered previously untested code paths, including unit tests for core methods, validation for operator logic, and scenarios for error handling in SQL generation. This contribution demonstrates that agents can improve test suites not only by increasing the overall coverage percentage, but also by adding targeted tests for specific areas like individual methods and critical error handling scenarios.

\smallskip
\textbf{\textit{Code refinement and optimization make up 12.7\% of \APRs, compared to just 3.2\% of \HPRs.}} According to Table~\ref{table:label_centered}, when combining style changes (\styleAPR{} vs. \styleHPR), and performance enhancements (\perfAPR{} vs. \perfHPR) are consistently more prevalent in \APRs. These changes often target repetitive, rule-based tasks, for example, a PR eliminating wildcard column selections to optimize database queries.\footnote{\url{https://github.com/mattermost/mattermost/pull/30424}} Similarly, stylistic cleanups, like resolving linter warnings and refining naming conventions,\footnote{\url{https://github.com/Nayshins/ummon/pull/32}} exemplify areas where AI reduces human workload. This pattern suggests that developers leverage agents to automate such tedious, rule-based work, reducing human workload on tasks related to code conventions and micro-optimizations.

\smallskip
\textbf{\textit{Documentation updates are more frequent in \APRs~(\printFix{\calcDocsAPR}\%) than in \HPRs~(\printFix{\calcDocsHPR}\%), indicating significant AI involvement in maintaining textual artifacts.}}
As shown in Table~\ref{table:label_centered}, documentation changes were observed in \docsAPR{} out of \purposeTotalAPR{} \APRs and \docsHPR{} out of \purposeTotalHPR{} \HPRs. These changes include updates to user-facing documentation, code comments, and onboarding materials. For example, an \APR enhanced project documentation by adding practical code examples in docstrings, correcting formatting inconsistencies, and refining configuration descriptions to improve clarity for new users.\footnote{\url{https://github.com/brentyi/tyro/pull/266}}

\smallskip
\textbf{\textit{Agents and humans contribute almost equally to maintenance and configuration tasks, accounting for 20.7\% and 21.2\% of pull requests, respectively.}} 
According to Table~\ref{table:label_centered} when combining the \emph{\textsc{build}}, \emph{\textsc{ci}}, and \emph{\textsc{chore}} categories, \APRs account for 20.7\% of these tasks, a figure very close to the 21.2\% from \HPRs. 
Agents demonstrate human-level performance on project-specific tasks, including dependency upgrades and workflow configuration. 
The ability of agents to manage complex, policy-driven work is well-exemplified by a PR\footnote{\url{https://github.com/shandley/awesome-virome/pull/43}} that enhanced CI workflows with stricter permissions and PR automation. Furthermore, their consistent assistance in routine build improvements, like updating compilation flags and frameworks\footnote{\url{https://github.com/giselles-ai/giselle/pull/687}} or performing dependency maintenance,\footnote{\url{https://github.com/dxos/dxos/pull/8814}} confirms that the role of agents in these areas is as significant as that of human developers.

\textbf{\textit{Multi-purpose PRs are more common in \APRs, suggesting agents frequently target overlapping objectives within a single PR.}} Among 213 \APRs, 85 \APRs~(39.9\%) have multiple objectives, compared to just 27 multi-purpose \HPRs~(12.2\%). This reflects an agentic coding pattern where AI consolidates repetitive or complementary tasks into a single PR.
A detailed analysis of label combinations reveals that the most common pairings include \emph{\textsc{feature development}} combined with \textsc{test} (9.0\%), \emph{\textsc{refactoring}} with \emph{\textsc{test}} (7.7\%), and \emph{\textsc{bug fixes}} with \emph{\textsc{test}} (7.7\%). 
These results suggest that agents have a consistent development of simultaneously creating and updating relevant test code when performing primary coding tasks (feature development, refactoring, and bug fixing). This allows for both code changes and their quality assurance to be achieved simultaneously within a single PR.
One such example\footnote{\url{https://github.com/github/github-mcp-server/pull/118}} updated an SDK to use a new, more powerful API endpoint (feature development). At the same time, the agent meticulously updated the corresponding test suite to validate the new functionality (test). This pattern of integrating development with immediate quality assurance showcases how agents can produce more robust and reliable changes within a single, cohesive contribution.

\begin{figure*}[t]
\centering
\includegraphics[width=0.9\linewidth]{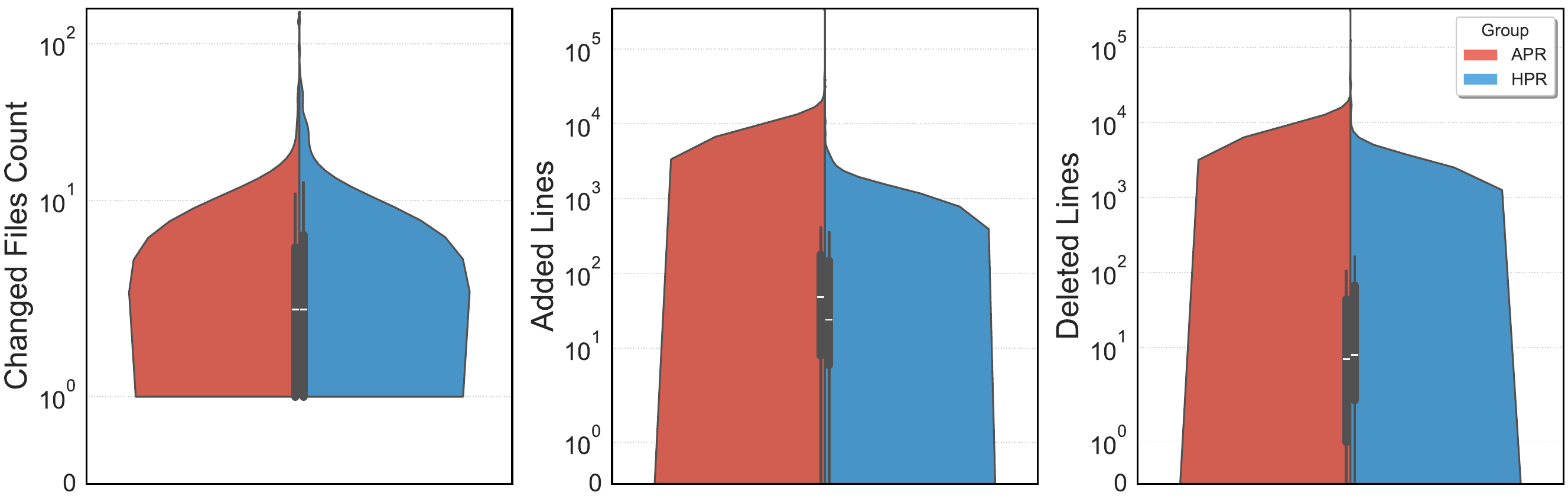}
\caption{Distribution of change metrics including changed files, added lines, and deleted lines in revised commits. Note that these do not include metrics from the first commit. }
\label{fig:measurement-initial}
\end{figure*}

\smallskip
\textbf{\APRs introduce more code and include longer descriptions than \HPRs.} \fig{fig:measurement-initial} shows the distributions of these metrics. Examining initial submissions, both \APRs and \HPRs modify a median of \firstchangeFileAPR\xspace files.
\aprs add more lines (\firstAddedLineAPR vs \firstAddedLineHPR) while maintaining similar deletion counts (\firstDeletedLineAPR vs \firstDeletedLineHPR). Moreover, \aprs contain significantly longer PR descriptions with a median of \firstTextAPR words, compared to \firstTextHPR words in \hprs.
Mann-Whitney U tests ($\alpha = 0.05$) confirm that the differences description length are statistically significant. One reason behind this could be that the agent is documenting its process, its reasoning, and the changes it made in detail. This could be helpful for reviewers.

\summarybox{\textbf{Answer to RQ1}}{
Both \APRs and \HPRs primarily address bug fixes and feature development. \APRs more frequently involve refactoring, documentation updates, and test improvements, while \HPRs focus relatively more on project-specific maintenance tasks such as CI and dependency management. Additionally, \APRs are more likely to serve multiple purposes in a single submission, as well as include more code additions and longer PR descriptions.
}

\bigskip
\subsection{\rqB}\label{sec:rqb}

\approach
We examine the proportion of \APRs~(\aprs) and \HPRs~(\hprs) that are eventually merged. In this study, we consider merged PRs as accepted and closed-but-unmerged PRs as rejected. We compute the merge rate as the ratio of merged PRs to the total number of \aprs and \hprs (\ie including both merged and closed PRs). We also analyse the time to merge and conduct a log-rank test to evaluate the statistical significance of differences in merge times. In addition, we investigate the reasons for PR rejections by manually inspecting 92 rejected PRs.

To categorize the reasons for PR rejections, we adopted the classification framework proposed by Pantiuchina\etal\cite{DBLP:journals/tosem/PantiuchinaLZPL22}, which identifies rejection patterns in refactoring PRs through empirical analysis of code review comments. Although their work focused on refactoring, the framework applies to our broader context, as many rejection reasons are independent of the change type.
The framework systematically organizes rejection reasons into process-related and refactoring-specific categories. We use five subcategories: (i) three from process-related subcategories called \emph{\textsc{Are inactive (author/community)}}, \emph{\textsc{Are too large}}, and \emph{\textsc{Are obsolete}} ; and (ii) two from refactoring-specific subcategories called \emph{\textsc{Do not add value}} and \emph{\textsc{Contain choices of non-optimal design solutions}}. 
The classification process by two independent authors achieved 68.5\% agreement between the two primary annotators, with a Cohen's K coefficient of 0.4865, indicating moderate agreement~\cite{kappa}.

The selection of these categories was motivated by two factors: First, preliminary analysis of our dataset revealed that these rejection patterns occurred with high frequency. Second, employing an empirically-validated classification enhances the reliability of our findings and facilitates comparison with existing literature.
However, during our analysis, we identified rejection reasons that could not be adequately classified within the framework from Pantiuchina\etal\cite{DBLP:journals/tosem/PantiuchinaLZPL22}. For these cases, we inductively derived new categories through open coding of rejection comments, thereby extending the original classification to comprehensively capture the rejection patterns specific to our context.

\begin{table}[b]
 \footnotesize
    \centering
    \caption{Acceptance and merge time statistics of PRs.}
    \begin{tabular}{crrrrr}\toprule
                &  \# Accepted & \# Rejected & \# Total& \% Acceptance & Median time to merge \\\midrule
        \aprs&  \mergedAPRs&\closedAPRs&\studiedAPRs&\mergeRateAPR\% & 1.23 hours\\
        \hprs& \mergedHPRs& \closedHPRs&\studiedHPRs&\mergeRateHPR\% & 1.04 hours\\ \bottomrule
    \end{tabular}
    \label{tab:my_label}
\end{table}
\par

\results
\textbf{\textit{\mergeRateAPR\% of \APRs are accepted by human reviewers, although their acceptance rate remains statistically lower than \HPRs (\mergeRateHPR\%).} } 
Table~\ref{tab:my_label} shows that \mergedAPRs{} out of \studiedAPRs{} \aprs were merged, compared to \mergedHPRs{} out of \studiedHPRs{} \hprs. A chi-squared test revealed that \HPRs have a statistically significantly ($p < 0.05$) higher acceptance rate than \APRs. However, a log-rank test shows no significant difference in merge times between \APRs and \HPRs, suggesting comparable review efficiency for AI-generated contributions. The median time to merge for \aprs is \survivalTimeAPR hours, while \hprs take \survivalTimeHPR hours.

\smallskip
\textbf{\textit{The most common reasons for \APRs rejections stem from project evolution and PR complexity, not just code quality.}}
Table~\ref{table:reject_reason} shows that \emph{\textsc{alternative solutions}}~(\printFix{\calcdevelopAtherWayAPR}\%), \emph{\textsc{obsolescence}}~(\printFix{\calcObsoleteAPR}\%), and \emph{\textsc{oversized PRs}}~(\printFix{\calcTooLargeAPR}\%) frequently result in \APRs rejections, reflecting project evolution and maintainability concerns. 
For example, one \APRs\footnote{\url{https://github.com/mattermost/mattermost/pull/30593}} was closed after the team resolved the issue using a different solution, stating, ``\textit{We might come back to this, but for now solving the underlying question in a different way.}'' Another PR\footnote{\url{https://github.com/OpenAdaptAI/OpenAdapt/pull/946}} became obsolete after a newer contribution addressed the same functionality. Similarly, a large  PR\footnote{\url{https://github.com/solvespace/solvespace/pull/1553}} was closed with the comment, ``\textit{Closing in favor of smaller, more focused PRs to make reviews more manageable},'' emphasizing the difficulty of integrating oversized contributions into collaborative review processes.

\smallskip
\textit{\textbf{Process-related issues such as verification-only submissions~(\printFix{\calcRejectTestAPR}\%) and merge conflicts~(\printFix{\calcConflictAPR}\%) also contribute to rejections in \APRs.}  }
Table~\ref{table:reject_reason} shows that \testPR{} rejected \aprs were created solely for automated checks, \conflict{} encountered costly merge conflicts. For example, one verification-only PR\footnote{\url{https://github.com/giselles-ai/giselle/pull/599}} was  closed after confirming intended CI processes. A different PR\footnote{\url{https://github.com/openai/codex/pull/612}} faced extensive merge conflicts the contributor could not resolve, prompting abandonment. 
These cases illustrate how agentic coding contributions can be rejected for logistical or procedural reasons common in development workflows, making this category the second most frequent cause.

\begin{table*}[t]
 \centering
 \scriptsize
 \caption{Reasons for \APRs rejection}
 \label{table:reject_reason}
 \begin{tabularx}{\textwidth}{@{} l X r @{}}
  \toprule
  Category & Description & \% \aprs \\
  \midrule
  Are implemented by other PRs/developers & The contributor or project team chooses a different solution before this PR can be merged
  & \printFix{\calcdevelopAtherWayAPR}\% \\
  \addlinespace
  Submission for verification & The PR is created solely to trigger automated checks (e.g., CI pipelines) and is not intended for merging
  & \printFix{\calcRejectTestAPR}\% \\
  \addlinespace
  Are too large     & The PR is too large or complex, making effective review impractical
  & \printFix{\calcTooLargeAPR}\% \\
  \addlinespace
  Are obsolete   & The proposed changes become outdated or irrelevant due to evolving project requirements or newer implementations
  & \printFix{\calcObsoleteAPR}\% \\
  \addlinespace
  Are inactive (author/community)  & A project's state of ceased or minimal development activity, often implying a lack of ongoing maintenance or community engagement
  & \printFix{\calcInactiveAPR}\% \\
  \addlinespace
  Contain choices of non-optimal design solutions       & The PR implements a design that is considered suboptimal, inefficient, or architecturally unsound
  & \printFix{\calcNonOptimalAPR}\% \\
  \addlinespace
  Do not add value     & The PR provides no clear or significant benefit to the project, its users, or its maintainers
  & \printFix{\calcNotAddValueAPR}\% \\
  \addlinespace
  Increase complexity & The proposed solution was more complex than warranted
  & \printFix{\calcComplexAPR}\% \\
  \addlinespace
  No confidence in AI-generated code  & The PR is rejected because the code was AI-generated and lacks sufficient human review or understanding to ensure reliability
  & \printFix{\calcLackOfTrustAPR}\% \\
  \addlinespace
  Introduce bugs/ break APIs / breaks compatibility & The PR introduces bugs into existing functionality, breaks API compatibility, or includes changes that disrupt normal system operation 
  & \printFix{\calcBugsAPR}\% \\
  \addlinespace
  Are not in the community interest & The PR does not align with the project's direction or community goals, and is judged not worth the investment of limited resources \black 
  & \printFix{\calcNotInterestAPR}\% \\
  \addlinespace
  Introduce merge conflicts  & Resolving the merge conflicts requires significant manual effort, beyond a simple rebase
  & \printFix{\calcConflictAPR}\% \\
  \addlinespace
  Not sure & The PR has review comments, but they are ambiguous or open to multiple interpretations, making it difficult to classify the specific rejection reason
  & \printFix{\calcNotSureAPR}\% \\
  \addlinespace
  Unknown (No feedback provided) & The PR was closed without explanatory comments or discussion, preventing classification of the actual rejection reason
  & \printFix{\calcNotDiscussRejectAPR}\% \\
  \bottomrule
 \end{tabularx}
\end{table*}

 
\smallskip
\textbf{\textit{Technical shortcomings, such as issues with code implementation quality rather than strategic choices, account for 4.4\% of \APRs rejections, with \printFix{\calcNonOptimalAPR}\%  rejected for non-optimal design solutions, \printFix{\calcComplexAPR}\% for overly complex implementations, and \printFix{\calcBugsAPR}\% for introducing bugs or breaking functionality.}}
Table~\ref{table:reject_reason} shows that 4.4\% of the rejected \APRs are due to technical reasons where the implementation approach or code quality was problematic.
For example, one PR\footnote{\url{https://github.com/zenml-io/zenml/pull/3375}} was closed for bypassing project-specific serialization mechanisms, violating architectural principles. Another PR\footnote{\url{https://github.com/pytorch/test-infra/pull/6410}} was rejected after reviewers found that existing themes could achieve the same functionality with reduced complexity.
Furthermore, a PR\footnote{\url{https://github.com/pytorch/test-infra/pull/6400}} closed with the comment ``\textit{this is badly extracted and has some hallucinations, new CSS changes, and more importantly breaks the UI},'' demonstrating functional failures. These examples highlight persistent limitations in AI's ability to generate quality code that maintains system integrity.

\smallskip
\textbf{\textit{Strategic misalignment, such as issues with what to build rather than how to build it, accounts for 2.2\% of \APRs rejections, with \printFix{\calcNotAddValueAPR}\% closed for not adding value and \printFix{\calcNotInterestAPR}\% for not aligning with community interests. }}
Table~\ref{table:reject_reason} indicates these PRs targeted wrong problems or proposed unnecessary changes, regardless of implementation quality. 
For instance, one PR\footnote{\url{https://github.com/mattermost/mattermost/pull/30890}} was closed after maintainers determined the changes, while technically correct, failed to address the actual performance bottlenecks that mattered. Another PR\footnote{\url{https://github.com/MaiM-with-u/MaiBot/pull/195}} was rejected after reviewers referenced the project's policy document, determining the proposed large-scale changes did not provide sufficient benefit to justify their inclusion. These cases demonstrate challenges in AI's understanding of project priorities and identifying which problems actually need solving.

\smallskip
\textbf{\textit{Contributor inaction accounts for \printFix{\calcInactiveAPR}\% of \APRs rejections, with PRs abandoned when contributors stop responding to the review process.}}  
Table~\ref{table:reject_reason} indicates that \inactive{} rejected \aprs were closed due to general inactivity. For instance, a PR\footnote{\url{https://github.com/osmosis-labs/osmosis/pull/9029}} was marked stale and automatically closed after the contributor failed to respond to implementation concerns. Another case\footnote{\url{https://github.com/ApeWorX/ape/pull/2532}} involved a reviewer commenting ``Waiting on this to be resolved'' with a link to specific code review feedback, but the contributor never addressed the issue, resulting in automated closure by a bot. These examples illustrate how \APRs, when not actively managed by human contributors, face higher rejection risks due to minimal post-submission engagement.

\smallskip
\textbf{\textit{Trust remains a barrier for \APRs, with \printFix{\calcLackOfTrustAPR}\% of rejected \aprs explicitly closed due to lack of confidence in their correctness.}} 
As shown in Table~\ref{table:reject_reason}, \lackOfTrust{} rejected \apr was withdrawn because contributors or maintainers doubted the reliability of AI-generated code. For instance, one contributor closed their PR\footnote{\url{https://github.com/beekeeper-studio/beekeeper-studio/pull/2962}}, stating explicit concerns about validating Claude's output. This reflects a key socio-technical barrier, where agent-assisted development outpaces human trust in AI capabilities, limiting the integration of such contributions despite technical potential.

\smallskip
\textbf{\textit{The majority~(\printFix{\calcNotDiscussRejectAPR}\%) of rejected \APRs were closed without explanatory comments or discussion.}}
Table~\ref{table:reject_reason} shows that \notDiscuss{} out of \rejectTotalAPR{} rejected \aprs fell into this category. For instance, a PR\footnote{\url{https://github.com/anthropics/anthropic-cookbook/pull/145}} was closed without discussion or review comments, leaving the rejection reasons entirely unknown. This lack of feedback highlights a transparency challenge when evaluating why AI-generated contributions are declined.

\summarybox{\textbf{Answer to RQ2}}{
\APRs are widely accepted in real-world projects, with a \mergeRateAPR\% acceptance rate and similar review times to \HPRs. However, their acceptance is lower than human-written PRs (\mergeRateHPR\%). Most rejections stem from project context, such as the changes being implemented by other developers, submissions made purely for verification, or the PR's size, rather than inherent flaws in AI-generated code. Still, technical issues (e.g., suboptimal design, complexity) and trust concerns persist. 
}
\smallskip

  

\bigskip
\subsection{\rqC}\label{sec:rqc}
\approach
To understand the extent to which developers revise \APRs, we conducted a two-step analysis. First, to identify the proportion of agent-generated changes accepted without any modification, we calculated the percentage of \APRs and \HPRs that were merged with a single commit, containing only the initial submission. This metric allows us to quantify the rate at which agent-assisted changes are deemed sufficient without further developer intervention. Second, for the subset of \aprs that underwent at least one revision, we analysed the subsequent revision effort required to get the PR merged. We quantified this effort using the following four metrics, measured from the second commit to the final commit in the PR: (1) the number of revision commits, (2) the number of files changed, (3) the total number of lines added/deleted, (4) the percentage of code modification compared to the initial submission. This analysis provides insight into the cost and nature of developers' modifications to agent-generated code.

\begin{figure}[t]
\centering
\includegraphics[width=0.4\linewidth]{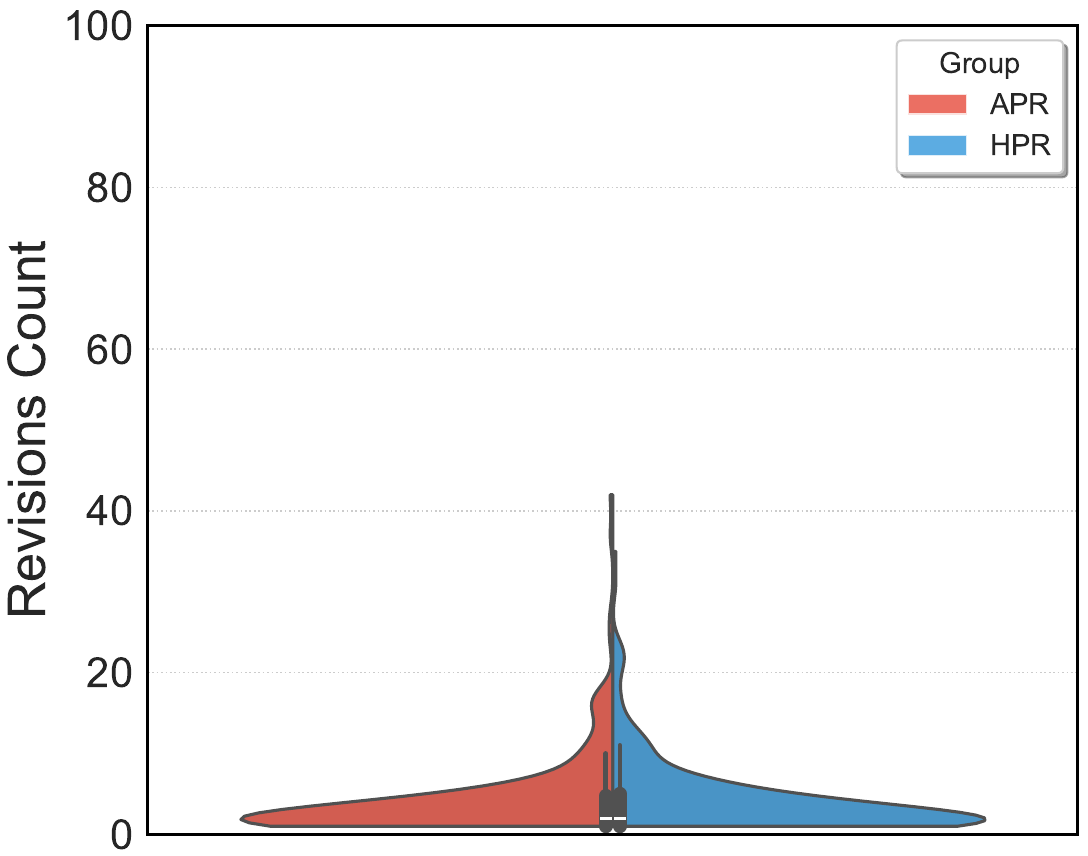}
\caption{The number of revisions commits in \APRs and \HPRs. The inner box shows the interquartile range, and the white lines indicate the medians.}
\label{fig:commits}
\end{figure}

\results
\textbf{\textit{The majority of both \APRs and \HPRs are merged without revision.}} Among merged PRs, we observe that \directMergeRateAPR\% of \aprs (\directMergedAPRs{} PRs) are merged as-is with a single commit, as compared to \directMergeRateHPRs\% (\directMergedHPRs{} PRs) of \hprs. In addition, the chi-squared test revealed no statistically significant difference between the two groups (p > 0.05). This similarity indicates that, in many cases, changes proposed by agentic coding are accepted without further human intervention, suggesting a baseline level of trust and adequacy in the initial agent-generated code.

\begin{figure*}[t]
\centering
\includegraphics[width=0.95\linewidth]{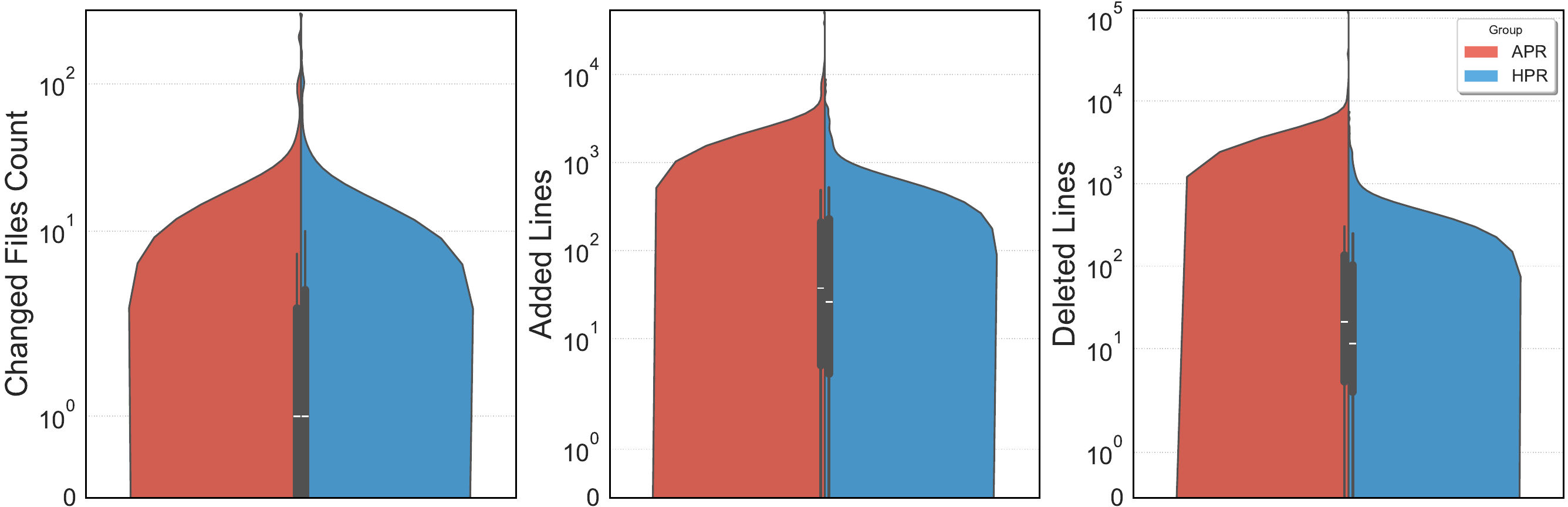}
\caption{Distribution of change metrics including changed files, added lines, and deleted lines in revised commits. Note that these do not include metrics from the first commit. }
\label{fig:measurement}
\end{figure*}

\smallskip
\textbf{\textit{When revisions are required, there is no statistically significant difference in revision frequency or cost.}}
Analysis of revision patterns shows that both \aprs and \hprs typically include a median of \revisionsAPR{} revised commits prior to merge (see \fig{fig:commits}). The Mann-Whitney U tests show no statistically significant differences between \APRs and \HPRs in terms of revision frequency. Although the median number of files, lines added, and lines deleted in revised commits can differ between the two groups (see \fig{fig:measurement}), none of these differences are statistically significant at $\alpha = 0.05$.
These results suggest that the revision cost for \APRs is not significantly different from that for \HPR. Given this equivalence, leveraging AI to automate the initial stage of PR creation could effectively reduce developer burden and improve productivity.


\smallskip
When comparing revisions against the baseline (\ie the first submission), the additionally changed files represented a median increase of \changefilePercentageAPR\% for both \aprs and \hprs.
For changed lines, \hprs showed larger increases with a median of \printFix\changelinePercentageHPR\% compared to \printFix{\changelinePercentageAPR}\% for \aprs.\footnote{Some pull requests contain zero added or deleted lines in their initial submission, making it impossible to compute relative metrics based on additions or deletions alone. Therefore, we use the total number of modified lines, defined as the sum of added and deleted lines.} 
This suggests that the revision patterns between AI-generated and human-written PRs are statistically similar. 

\summarybox{\textbf{Answer to RQ3}}{
The proportion of \APRs and \HPRs that did not require revisions was similar, at \directMergeRateAPR\% and \directMergeRateHPRs\%, respectively. Furthermore, a Mann-Whitney U test on the revision effort measured by the number of commits, lines of code added/deleted, and files modified revealed no statistically significant difference between 
\APRs and \HPRs.
}
\bigskip
\subsection{\rqD}\label{sec:rqd}

\approach
We examined all \mergedAPRs\xspace merged \APRs and identified \notDirectolyMergedAPRs\xspace \APRs~(\printFix{\calcNotDirectolyMergedAPRs}\%) that had modifications between initial submission and merge. We performed manual classification on these \notDirectolyMergedAPRs\xspace \APRs to understand what types of changes were made during the review process. We manually inspect all commits added after the initial submission and classify them using the same two-dimensional framework from Zeng\etal\cite{DBLP:conf/icse/ZengZQL25} described in Section~\ref{sec:rqa}. For each PR, we aggregated revision types at the PR level to avoid over-representing PRs with numerous small commits. For example, if a PR included three bug fixes and two stylistic changes across separate commits, it was counted as one instance of each change type. 
This manual classification was performed by the second and third authors. As in Section~\ref{sec:rqa}, we treated this as a multi-label task and measured agreement at the label level. The overall agreement rate was 64.1\% (541 out of 844 labels). Disagreements were resolved by the second author following the same process described in Section~\ref{sec:rqa}.
In addition, to understand the extent of continued AI assistance during the revision process, we analyzed co-authorship patterns in revision commits. We extracted commit metadata to identify commits containing ``\texttt{Co-Authored-By: Claude}'' attribution.

\begin{table*}[t]
 \centering
 \scriptsize
 \caption{Revision based on analysis of PR content}
 \label{table:revision_label}
 \begin{tabularx}{\textwidth}{@{} l X r @{}}
  \toprule
  Category & Description & \% \aprs \\
  \midrule
  fix      & Code changes that fix bugs and faults within the codebase 
  &  \printFix{\calcFixRevision}\% \\
  \addlinespace
  docs     & Updates to documentation or comments, such as README edits, typo fixes, or API docs improvements 
  & \printFix{\calcDocsRevision}\% \\
  \addlinespace
  refactor & Code restructuring without changing its behavior, aiming to improve maintainability 
  & \printFix{\calcRefactorRevision}\% \\
  \addlinespace
  style    & Non-functional code changes that improve readability or consistency. This type encompasses aspects like variable naming, indentation, and addressing linting or code analysis warnings 
  & \printFix{\calcStyleRevision}\% \\
  \addlinespace
  chore    & Project‐wide housekeeping tasks such as dependency bumps, version increments, and miscellaneous cleanup 
  & \printFix{\calcChoreRevision}\% \\
  \addlinespace
  test     & Additions or modifications to test files, including new test cases or updates to existing tests 
  & \printFix{\calcTestRevision}\% \\
  \addlinespace
  feat     & Code changes that introduce new features to the codebase, encompassing both internal and user-oriented features 
  & \printFix{\calcFeatRevision}\% \\
  \addlinespace 
  build    & Changes to build configurations (e.g., Maven, Gradle, Cargo). Change examples include updating dependencies, configuring build configurations, and adding scripts 
  & \printFix{\calcBuildRevision}\% \\
  \addlinespace
  ci       & Changes to CI/CD workflows and configurations, e.g., ``.travis.yml'' and ``.github/workflows'' 
  & \printFix{\calcCiRevision}\% \\
  \addlinespace
  perf     & Code changes that improve performance, such as enhancing execution speed or reducing memory consumption 
  & \printFix{\calcPerfRevision}\% \\
  \bottomrule
 \end{tabularx}
\end{table*}

\results
\textbf{\textit{The majority~(\printFix{\calcFixRevision}\%) of revisions to \APRs target bug fixes.}}
As shown in Table~\ref{table:revision_label}, the most common revision type was fixing functional bugs, accounting for \printFix{\calcFixRevision}\%~(\fixRevision\xspace out of \revisionTotalAPR) of revised \APRs.  These fixes typically involved critical error handling improvements, and representative revisions addressed situations where critical errors were improperly handled. In one case,\footnote{\url{https://github.com/qltysh/qlty/pull/1591/commits/87444f54e702a64f519ca1052380bf9f50f7be96}} concurrent error propagation required introducing channel-based communication mechanisms to ensure fatal errors were immediately surfaced from worker closures. Similarly, another revision\footnote{\url{https://github.com/qltysh/qlty/pull/1588/commits/a38e75cc15657d8864cff66b388189b1ea677e19}} upgraded file operation failures from warnings to fatal errors, recognizing that certain failure conditions should halt execution rather than proceed with a potentially corrupted state. These patterns reveal that agent-generated code frequently implements optimistic error handling strategies that fail to distinguish between recoverable and non-recoverable failure conditions.

\smallskip
\textbf{\textit{Addressing documentation gaps constitutes the second largest focus (\printFix{\calcDocsRevision}\%) of \APRs revisions.}}
Documentation updates were present in \printFix{\calcDocsRevision}\%~(\docsRevision{} out of \revisionTotalAPR{}) of revisions, as shown in Table~\ref{table:revision_label}. Although agents sometimes generated or updated code comments, they often failed to synchronize all relevant artifacts. For example, one revision\footnote{\url{https://github.com/brentyi/tyro/pull/266/commits/8d3bf92a48a8bef4dbb09bfc31bc39dbfb544046}} removed an outdated optional dependencies section from installation documentation that should have been deleted alongside related code changes. This disconnect means that, in practice, human reviewers spend a non-trivial amount of effort ensuring that documentation, README files, and code comments accurately reflect code changes from agents.

\smallskip
\textbf{\textit{Refactoring is frequently required~(\printFix{\calcRefactorRevision}\%) to improve code structure before merging \APRs.}}
Table~\ref{table:revision_label} shows that \refactorRevision\xspace out of \revisionTotalAPR\xspace modified \APRs involved refactoring, highlighting the need for structural improvements in these submissions. These structural changes, which do not affect program behaviour, were frequently necessary to eliminate code duplication, enhance modularity, or revise architectural decisions made by the agent. For example, a reviewer consolidated redundant initialization logic spread across multiple entry points into a unified service.\footnote{\url{https://github.com/basicmachines-co/basic-memory/pull/88/commits/ca5fa557c25bead9aafcf545544957d835bba798}} In this case, common functionality for database migrations and file synchronization was extracted from separate components and moved into a shared module, improving both error handling and logging consistency. This finding suggests that although the initial submissions from \APRs were functionally correct, reviewers often had to refactor them to align better with project architecture, enhance maintainability, and reduce technical debt.

\smallskip
\textbf{\textit{A common portion~(\printFix{\calcStyleRevision}\%) of revisions is commonly focused on polishing code style rather than behaviour.}}
As shown in Table~\ref{table:revision_label}, \styleRevision\xspace out of \revisionTotalAPR\xspace modified \APRs received style-related revisions during review, indicating code quality issues in initial submissions. Typical fixes include enforcing naming conventions, correcting formatting, and resolving linter warnings that agentic tools missed. For example, a reviewer addressed multiple static analysis violations that were present in the agent's original code,\footnote{\url{https://github.com/INCATools/ontology-access-kit/pull/831/commits/5a6f71986fcfaafa291b9aa405d51d76c8c27c97}} including unused imports, exception variables that were declared but never referenced, and incorrect import ordering. 
These changes are largely cosmetic but necessary for integration, and suggest that current agents often underperform on project-specific style rules, requiring human intervention to maintain readability and consistency. Static analysis violations and ignored best practices were common triggers for these types of revisions.

\smallskip
\textbf{\textit{Project Housekeeping tasks account for \printFix{\calcChoreRevision}\% of revisions.}}
According to Table~\ref{table:revision_label}, chores represent \printFix{\calcChoreRevision}\% (\choreRevision{} out of \revisionTotalAPR\xspace) of revisions, typically involving project-wide metadata such as version bumps or release notes that the agent overlooked. For example, one revision\footnote{\url{https://github.com/yumemi-inc/Tart/pull/106/commits/88594d8a116e7c91390740cf7e9ddb7ccfaa61f5}} involved a simple but critical version bump from ``3.0.0-alpha01'' to ``3.0.0-alpha02'' in a Gradle configuration file (\texttt{libs.versions.toml}), a necessary step for a new release that the agent did not perform. This pattern indicates a specific gap in agent capabilities: while they successfully modify application code, they often fail to propagate corresponding changes to project-level configuration files. Consequently, these essential administrative tasks fall to human reviewers to ensure the project's versioning and release process remains consistent and accurate.

\smallskip
\textbf{\textit{New features are added in \printFix{\calcFeatRevision}\% of \APRs by human reviewers.}}
Table \ref{table:revision_label} indicates that \featRevision\xspace out of \revisionTotalAPR\xspace modified \APRs (\printFix{\calcFeatRevision}\%) received feature additions during the review process. For example, one revision\footnote{\url{https://github.com/github/github-mcp-server/pull/118/commits/0527bc552591df807ee7be0b3a79f428d2f0ed2c}} expanded the PR review API functionality that the agent had initially implemented with basic capabilities. The revision added support for multi-line comments, including new parameters for line positioning, validation logic for parameter combinations, and comprehensive test coverage. This enhancement demonstrates that agent-generated implementations often provide core functionality but miss advanced features or edge cases that developers identify during review, requiring subsequent additions to meet complete user requirements.

\smallskip
\textbf{\textit{Build configuration adjustments occur in \printFix{\calcBuildRevision}\% of reviewed \APRs.}}
Table~\ref{table:revision_label} shows that \buildRevision\xspace out of \revisionTotalAPR\xspace modified \APRs received build configuration updates during review. 
One revision\footnote{\url{https://github.com/coder/coder/pull/17035/commits/3c3aa219b895a981dd13a1b530603aa257f5549e}} addressed compatibility issues between the linting tool and the newer Go version. 
The agent's initial implementation did not account for version incompatibilities, causing build failures with Go 1.24.1. The revision updated the linter version, configured it to run with a reduced set of checks, and modified the build process to continue despite linting failures. This temporary solution allowed the build pipeline to function while a comprehensive fix was planned for a subsequent update. The revision illustrates that agent-generated build configurations may overlook toolchain compatibility requirements, 
requiring manual intervention to maintain functional build processes across version updates.

\smallskip
\textbf{\textit{Human reviewers also shore up test coverage that agents frequently omit.}} Although tests account for only \printFix{\calcTestRevision}\%~(\testRevision{} out of \revisionTotalAPR{}) of the revisions in Table~\ref{table:revision_label}, the added suites are often substantial and guard edge cases and failure paths that the original submission did not address. For instance, one revision\footnote{\url{https://github.com/aleph-im/aleph-vm/pull/786/commits/bf69485b01c6d246126bd471f5f1509eef4cc84d}} added comprehensive unit tests for GPU X-VGA support detection functionality that were absent from the agent's initial submission. The new tests cover device-argument generation, detection-process validation, error-handling scenarios, and configuration-flow verification. The agent had implemented the GPU support feature without corresponding test coverage, which reviewers identified as a gap requiring remediation. This pattern echoes our earlier finding in Section~\ref{sec:rqb} that reviewers explicitly reject \APRs when they cannot fully trust the code without further human verification.

\smallskip
\textbf{\textit{CI/CD and performance optimizations represent a smaller but crucial tail of revisions (8.4\%).}} CI/CD modifications are relatively rare, comprising \printFix{\calcCiRevision}\%  of revisions, yet they are essential for maintaining a healthy pipeline. One revision, for instance, introduced a script to enforce version-string consistency across build artifacts after the original agent submission broke continuous integration.\footnote{\url{https://github.com/coder/coder/pull/17164/commits/de29681960767cb7eb01b407fa6de4fa48a07976}} Performance improvements appear in an even smaller fraction, just \printFix{\calcPerfRevision}\%, such as when a reviewer added a cache-busting mechanism to ensure data freshness for an efficient but potentially stale storage driver implemented by an agent, thereby achieving both accuracy and performance.\footnote{\url{https://github.com/giselles-ai/giselle/pull/695/commits/f9356079fe49ce56ca04e821098432916367032b}}
Taken together, these results highlight that while agentic coding already addresses the majority of functional and stylistic changes, human reviewers continue to play a key role in ensuring project hygiene, pipeline reliability, and targeted performance optimization.

\smallskip
\textbf{\textit{Agents remain actively involved in 41.1\% of all revisions for \APRs.}} Our analysis shows that 41.1\%~(88 out of \notDirectolyMergedAPRs{}) of the revised \APRs were co-authored with Claude, accounting for 34.1\%~(298 out of 873) of the commits made during the revision process. This indicates that developers frequently rely on AI tools not only for initial code generation but also for iterative refinement during review. The substantial proportion of AI co-authorship in revisions underscores the sustained role of agentic systems throughout the software development cycle.

\summarybox{\textbf{Answer to RQ4}}{
Among merged Agentic PRs, \printFix{\calcNotDirectolyMergedAPRs}\% require reviewer revisions, primarily for bug fixes (\printFix{\calcFixRevision}\%), documentation updates (\printFix{\calcDocsRevision}\%), refactoring (\printFix{\calcRefactorRevision}\%), and code style improvements (\printFix{\calcStyleRevision}\%).
This highlights that while AI-generated code is a strong starting point, human oversight is essential to ensure correctness, maintainability, and adherence to project standards.
}
\bigskip

\section{Implications}\label{sec:implications}

In this section, we discuss the implications of our findings for researchers, developers, and coding agent builders.

\subsection{Implications for researchers}

\textbf{\textit{Quantify the socio-technical cost of building trust in \APRs.}}
Although only 2.2\% of rejected \APRs explicitly cited ``lack of confidence,'' many rejections followed slow reviews or repeated revision requests~(Section~\ref{sec:rqb}), suggesting that reviewer skepticism toward AI-generated code operates more subtly than direct statements might indicate. This hidden friction represents a significant barrier to AI tool adoption that traditional metrics fail to capture. Mixed-method studies that combine review-time analytics with developer interviews could reveal how latent scepticism increases cognitive load, and whether transparency cues (\eg provenance statements) mitigate this effect. In addition, given that nearly one-third of merged \APRs still required bug fixes, style adjustments, or refactoring (Sections~\ref{sec:rqc} and \ref{sec:rqd}), measuring discussion length, reviewer effort, and escaped-defect rates will help identify when AI contributions save work versus when they add overhead.

\smallskip
\textbf{\textit{Create PR-centric benchmarks that reward alignment across code, documentation, and review resolution.}}
Current benchmarks such as SWE-bench~\cite{DBLP:conf/iclr/JimenezYWYPPN24} overlook documentation drift, style adherence, and reviewer negotiation, all of which accounted for 52.4\% of \APR revisions (Section~\ref{sec:rqd}). This gap between benchmark evaluation and real-world requirements represents a fundamental misalignment in how we assess coding agent capabilities. While existing benchmarks focus predominantly on functional correctness, whether the generated code passes test cases, they fail to capture nearly half of the actual work required to integrate AI-generated code into production systems. 
We suggest building datasets that incorporate time-to-merge, rework size, and the alignment between documentation and implementation. Such benchmarks can be applied during both training and evaluation of coding agents to promote improvements beyond functional correctness.

\smallskip
\textbf{\textit{Instrument the entire human-AI workflow, not isolated tools.}}
Our results show that 34.1\% of revision commits were co-authored by Claude Code~(Section~\ref{sec:rqd}), and review threads frequently referenced GitHub Copilot and other agents. This widespread multi-agent collaboration highlights a critical gap in our understanding of agent-assisted development workflows. We recommend that future studies collect comprehensive full-trace datasets that capture the entire development lifecycle from initial prompts and agent planning phases to intermediate code generations, CI/CD logs, review iterations, and final merge decisions. Such datasets would reveal crucial interaction patterns, including where hand-offs between agents and developers occur, where duplicate efforts waste resources, and where conflicting edits from different agents inflate latency and introduce defects. 

\subsection{Implications for developers}

\smallskip
\textbf{\textit{Split large, multi-purpose PRs into smaller, easy-to-review submissions.}}
Nearly 40\% of \APRs combined multiple tasks (Section~\ref{sec:rqa}), and ``too large'' was among the top three rejection reasons (Section~\ref{sec:rqb}). Large PRs increase reviewer cognitive load and raise the risk of rejection, as previous studies \cite{DBLP:conf/msr/WeissgerberND08} have warned developers for decades. This challenge remains relevant even for agents. In fact, RQ1 shows agents generate comprehensive solutions that attempt to address multiple issues simultaneously. 
The tendency of AI coding agents to produce extensive changes when given broad directives mirrors the problematic patterns seen in human development, but with potentially greater scale and complexity. 
When assigning a task to a coding agent, developers should adhere to the principle of providing a small, self-contained task that can be addressed in a single PR, or structure broader tasks to produce a sequence of smaller, follow-up PRs for wide-reaching changes. 

\smallskip
\textbf{\textit{Embed project-specific style, rules, and architecture in agent instructions.}}
Style mismatches accounted for \printFix{\calcStyleRevision}\% of revisions, while refactors accounted for another \printFix{\calcRefactorRevision}\% (Section~\ref{sec:rqd}). Missing documentation and tests contributed \printFix{\calcDocsRevision}\% and \printFix{\calcTestRevision}\%, respectively. 
These statistics reveal a critical pattern: much of the revision work on AI-generated code results not from functional errors but from integration friction, which is the gap between what agents produce and what teams expect. This misalignment creates substantial hidden costs, as developers must repeatedly guide agents toward project-specific conventions that could have been specified upfront.
We recommend that developers maintain a dedicated guideline file (\eg \texttt{CLAUDE.md}, supported by Claude Code) that includes formatting rules, design principles, and architectural constraints. This practice can help coding agents produce outputs aligned with project standards, ensuring that code, documentation, and tests remain synchronized as part of the definition of done.

\subsection{Implications for coding agent builders}

\smallskip
\textbf{\textit{Coding agent builders should present uncertainty and supply review scaffolding, not just code.}}
RQ2~(Section~\ref{sec:rqb}) shows that PRs rejected for non-optimal design often lacked implementation rationale. This absence of reasoning transparency creates a critical bottleneck in the review process, as reviewers must expend significant effort inferring why certain design decisions were made, or worse, may reject sound solutions simply because the underlying logic remains opaque. We recommend attaching a ``confidence card'' to each PR, including the plan followed, key assumptions, considered alternatives, and known edge cases. A reviewer checklist could guide targeted human scrutiny.

\smallskip
\textbf{\textit{Automate low-risk PR maintenance tasks such as rebases, conflict resolution, and stale-response handling.} }
High-cost conflict resolution accounted for \printFix{\calcConflictAPR}\% of rejections (Section~\ref{sec:rqb}). These rejections often occur not because of fundamental code issues but due to timing misalignments. In other words, PRs that were valid when created become unmergeable as the target branch evolves. Currently, developers must manually monitor agent-generated PRs, performing routine maintenance tasks that consume time without adding meaningful value.
Agents could be enhanced to periodically rebase branches against the main branch, resolve simple conflicts, and automatically respond to staleness warnings. These capabilities would reduce manual oversight and help keep PRs mergeable.

\smallskip
\textbf{\textit{Coding agent builders should integrate additional tools to improve adherence to project conventions.}}
The high rates of style~(\printFix{\calcStyleRevision}\%) and refactoring~(\printFix{\calcRefactorRevision}\%) revisions indicate that generic models do not fully capture local conventions~(Section~\ref{sec:rqd}). 
This also implies that it creates persistent friction between agent capabilities and team expectations. This misalignment stems from agents being trained on diverse codebases with conflicting conventions, making them unable to infer project-specific patterns from limited context.
Using the Model Context Protocol (MCP) servers~\citep{hasan2025mcp}, builders can connect agents to linters (\eg \texttt{pylint}) and static analysis tools to enforce formatting and structural rules before submission. Similarly, build-checking and test-coverage tools could be integrated to catch common revision triggers in advance.

\section{Related Work}\label{sec:relatedwork}

\subsection{Human-AI collaboration in software engineering}

Recent research has emphasized the increasing collaboration between human developers and AI-powered technologies in common software engineering tasks~\cite{DBLP:journals/csur/LambiaseCPF25}, such as requirements documentation~\cite{DBLP:conf/icsm/RahmanZMRRS24}, code search~\cite{DBLP:conf/sigsoft/LiuZ0WH024}, and code generation~\cite{DBLP:journals/jss/DakhelMNKDJ23}. Hassan\etal~\cite{DBLP:journals/corr/abs-2410-06107} characterized this collaborative relationship within the emerging paradigm of Software Engineering 3.0, envisioning AI teammates that enhance developer productivity and reduce the cognitive load on human developers. 
However, despite these technical advances, human oversight remains crucial.
AI-powered tools have demonstrated potential for improving productivity, but human-in-the-loop approaches are still required to maintain high-quality results~\cite{DBLP:conf/aiware/CoutinhoMSDFS24, DBLP:conf/xpu/MockMR24}. Hence, establishing clear guidelines and best practices for managing human-AI interactions remains an essential step for industrial organizations to integrate these technologies~\cite{DBLP:conf/icst/SantosSMS24}.

In addition, multi-agent frameworks such as LLM-Agent, FlowGen and FightFire assign specialised roles (planner, coder, tester) to cooperating LLMs, emulating Agile practices and raising robustness~\cite{DBLP:journals/tosem/HeTL25,DBLP:conf/icse/LinKC25,DBLP:journals/tse/YuLHKLX24}. A recent systematic review consolidates prompt-engineering techniques and taxonomies of human-AI collaboration scenarios in software engineering~\cite{DBLP:journals/tosem/HouZLYWLLLGW24}.

However, despite these technical achievements, significant social and human challenges exist in AI-assisted development. Piorkowski\etal~\cite{DBLP:journals/pacmhci/PiorkowskiPWWMP21} and Nahar\etal~\cite{DBLP:conf/icse/NaharZLZK25} uncover persistent knowledge gaps inside multi-disciplinary AI teams, reinforcing the call for explicit interaction guidelines.
Furthermore, Adam\etal~\cite{10.1145/3721127} found that LLM-assisted reviews can disrupt collective accountability mechanisms inherent in traditional peer reviews, as developers value intrinsic factors such as professional integrity and reputation that cannot be maintained when interacting with AI systems. These studies highlight the complex socio-technical challenges in human-AI collaboration.


\subsection{AI-assisted issue reports and PRs}

Previous studies have explored AI-assisted activities related to issue reports~\cite{DBLP:conf/kbse/BoJ000024, DBLP:conf/icse/KangYY23} and PRs~\cite{DBLP:conf/msr/ChouchenBBOAM24, DBLP:conf/ease/WatanabeK0HYI24}. However, no prior research has investigated the use of agentic coding tools, such as OpenAI's Codex\footnote{\url{https://openai.com/index/introducing-codex}} and Claude Code, in real-world development contexts. 
Instead, current research primarily focuses on LLM-based chatbots like ChatGPT. 
For example, Chouchen\etal~\cite{DBLP:conf/msr/ChouchenBBOAM24} reported that ChatGPT is frequently used in review-intensive PRs to reduce developer effort. Despite this positive impact, 30.7\% of AI-generated suggestions during code reviews were still met with skepticism or disagreement~\cite{DBLP:conf/ease/WatanabeK0HYI24}. Our study fills this gap by tracing 567 autonomous PRs created by Claude Code across 157 open-source projects, reporting acceptance rates, revision effort, and rejection rationales.

Researchers have proposed various approaches for directly involving LLMs in activities related to issue reports and PRs. For example, Kang\etal\cite{DBLP:conf/icse/KangYY23} used LLMs to generate test cases from issue reports for easier bug reproduction, Irsan\etal\cite{DBLP:conf/icsm/IrsanZT0J22} trained LLMs to automatically generate PR titles, and Bo\etal\cite{DBLP:conf/kbse/BoJ000024} leveraged ChatGPT to improve the clarity of issue reports for helping subsequent tasks such as bug fixing. 

Complementary work investigates AI agents that clarify ambiguous requirements and collaboratively co-evolve PR descriptions; for instance, AgileGen aligns end-user perspectives with acceptance Zhang~\cite{DBLP:journals/tosem/ZhangXGXCZZFZ25}, while large-scale analyses of LLMs on code-change tasks reveal their potential for automated review of PR diffs~\cite{DBLP:journals/tosem/FanLLLXL25}. Multi-agent benchmarks also examine ChatGPT's ability to self-verify its generated code through test-report generation, reducing reviewer burden~\cite{DBLP:journals/tse/YuLHKLX24}.

Early work by Tufano\etal~\cite{DBLP:conf/icse/TufanoPTPB21} pioneered DL models that propose code changes before review submission and apply reviewer suggestions after submission. Guo\etal\ report that GPT-4 can refine 66\% of files given review comments~\cite{DBLP:conf/icse/GuoCXLL0024}, and later boost intent-extraction accuracy to 79\% in their ICSE '25 study~\cite{DBLP:conf/icse/GuoXL00B25}. To improve developer confidence, Di and Zhang~\cite{DBLP:conf/icse/Di025} interleave inline comments with LLM edits, cutting task time by 16.7 percent.


\subsection{Empirical studies of automatic code generation}
Researchers have studied various aspects of AI-generated code, including correctness~\cite{DBLP:conf/esem/BillahRCR24}, efficiency~\cite{DBLP:conf/forge/NiuZL0024}, and security~\cite{DBLP:journals/tse/LiuTLZZ24}. 
For example, Billah\etal~\cite{DBLP:conf/esem/BillahRCR24} evaluated the correctness (pass rate) of LLMs in solving programming challenges on competitive programming platforms such as LeetCode and Codeforces. They reported a decrease in the pass rate in live Codeforces contests compared to archived problems.
However, code correctness alone does not necessarily imply efficiency or optimality as code complexity increases, and extra effort is required to support AI models in generating efficient code~\cite{DBLP:conf/forge/NiuZL0024}.
From a security perspective, Liu\etal~\cite{DBLP:journals/tse/LiuTLZZ24} identified security vulnerabilities in AI-generated code and highlighted potential security risks that must be mitigated before deployment in production environments.
Beyond generating code at the granularity of individual functions or files, SWE-bench~\cite{DBLP:conf/iclr/JimenezYWYPPN24} was introduced to evaluate the performance of AI-powered code generation tools in resolving real-world GitHub issues.

Subsequent work has shown that ChatGPT often produces buggy or low-quality code that benefits from iterative repair~\cite{DBLP:journals/tse/LiuTLZZ24}, while self-collaboration frameworks with multiple LLM agents significantly improve correctness and reduce human intervention~\cite{DBLP:journals/tosem/DongJJL24}. Empirical studies highlight the challenges of domain-specific coding (\eg machine-learning pipelines) and demonstrate that knowledge-prompting strategies boost performance in such settings~\cite{DBLP:journals/tosem/GuCLHZWWXW25,DBLP:journals/tosem/ShinWWSW24}. Recent surveys map out future research directions for automatic programming pipelines, including integrated quality-assurance stages~\cite{DBLP:journals/tosem/LyuRRTT25}, and bias-testing frameworks reveal that LLMs can inherit societal biases, motivating prompt-engineering mitigation strategies~\cite{10.1145/3724117}.

Wang\etal~\cite{DBLP:conf/icse/WangZSHCMZ25} catalogue 557 distinct error patterns that six LLMs make on HumanEval. Niu\etal\ benchmark 19 pre-trained models across 13 tasks, confirming that functional correctness rarely correlates with efficiency~\cite{DBLP:conf/icse/NiuLNCGL23}. O'Brien\etal~\cite{DBLP:conf/icse/OBrienBIASR24} show Copilot frequently injects TODO-style self-admitted technical debt. For repair, Xia\etal~\cite{DBLP:conf/icse/XiaWZ23} and Bouzenia\etal\cite{DBLP:conf/icse/BouzeniaDP25} re-cast bug fixing as an autonomous planning problem executed by an LLM agent.



\section{Threats to Validity}\label{sec:threats_to_validity}

\subsection{Threats to internal validity}
A potential concern in our study is the ambiguity related to developers' behaviors when using Claude Code. Developers may modify the generated code before committing it. Conversely, even when developers use Claude Code to generate code, they may directly push the changes without requesting the tool to perform the commit itself.

To mitigate this threat, we explicitly included only PRs clearly marked as "Generated with Claude Code."
\blue
While this explicit labeling provides a strong signal of Claude Code's involvement and reduces potential ambiguity about code provenance, we acknowledge that it implies a binary classification that may overlook hybrid workflows. Regarding the Human baseline, the presence of undisclosed AI-generated code would likely dilute the differences between the two groups. Therefore, the significant differences observed in our study likely represent conservative estimates of the distinct characteristics of agentic workflows.
Furthermore, relying on brittle text markers (like ``Generated with Claude Code'') is insufficient for analyzing these hybrid workflows. Future research requires more fine-grained provenance tracking mechanisms to accurately separate agent-generated code from human modifications.

Another threat arises from the temporal discrepancy in data collection, which could introduce bias due to project evolution or seasonal factors. Regarding the baseline construction, we expanded the sampling window for \HPRs backward until the sample size matched that of \APRs, resulting in a collection period spanning approximately 10 months (back to April 2024). Our analysis of the temporal distribution shows that 42.0\% (238/567 PRs) of \HPRs were created within the exact same timeframe as \APRs (February 24 -- April 30, 2025). To address potential concerns about topical or seasonal drift, we conducted a sensitivity analysis comparing this time-aligned subset against the full baseline. We found that the distribution of task types in this subset remained consistent with Table~\ref{table:label_centered}. For example, ``fix'' and ``feat'' remained the dominant categories (27.0\% and 24.7\%, respectively), and the ratio of ``test'' (9.0\%) remained significantly lower than that of \APRs (18.8\%). Furthermore, performance metrics relevant to Table~\ref{tab:my_label} showed consistency; the subset achieved a comparable acceptance rate (87.0\% vs. 91.0\% for the full set) and median merge time (0.7h vs. 1.0h). These results suggest that our findings are robust and not significantly biased by the extended sampling window.
\black
\black
 
\subsection{Threats to construct validity} 
Claude Code implements agentic coding and can leverage the Model Context Protocol (MCP), which enables agents to integrate with external tools and data sources. However, this study does not isolate the specific contributions of agentic capabilities and MCP integration to the observed outcomes. Disentangling their individual impacts remains an important direction for future research.
\blue
Furthermore, our dataset excludes only open PRs to focus on contributions with definitive outcomes (merged or rejected). This exclusion creates a survival bias, as complex or controversial PRs that remain open for extended periods are not captured. Consequently, the reported acceptance rates and merge times reflect only completed workflows and may underestimate the latency of ongoing, undecided contributions.

Finally, our reliance on PR acceptance as a primary success metric does not account for long-term code quality or maintainability. A merged PR may still introduce latent bugs or negatively impact the codebase over time (\eg causing regressions or necessitating subsequent reverts). Since our study focuses on the immediate outcome of the review process, we lack longitudinal data to assess these downstream effects. Future research should investigate the long-term stability and maintainability of agent-generated code through longitudinal analysis.
\black

\subsection{Threats to external validity} 
This study examines only \studiedAPRs~PRs that explicitly used Claude Code, which represents the full set we were able to retrieve. One reason for this limited sample size is that Claude Code was launched only recently, in February 2025. 
\blue
Consequently, some observed developer behaviors may reflect an ``enthusiastic phase'' characterized by experimentation (\eg PR\footnote{https://github.com/thefrontside/effection/pull/992} explicitly noting ``I wanted to try Claude Code''). These patterns may differ from steady-state norms that emerge once the tool is fully integrated into standard workflows. Future work should expand this time window to determine if these patterns persist as the technology matures.

\black
To validate our findings, replication studies with larger datasets collected over a longer time frame will be necessary.
Finally, our analysis focused exclusively on Claude Code, an agentic coding tool whose usage could be reliably identified in PRs. As a result, our findings may not generalize to other agentic coding tools. 
\blue
We view this as a critical avenue for future research and encourage the community to investigate how different agentic coding tools (\eg GitHub Copilot) influence PR characteristics and acceptance rates.
\black

\section{Conclusion}\label{sec:conclusion}
This paper presents the first empirical study investigating the impact of agentic coding tools, specifically Claude Code, on open-source projects. We analyzed \studiedAPRs~pull requests across \projects~open-source projects to understand how these AI-generated contributions are received by developers. Our findings show that while \APRs are accepted at a lower rate than \HPRs~(\mergeRateAPR\% vs. \mergeRateHPR\%), they are still widely adopted in real-world projects. Importantly, when revisions are required, the extent of modifications does not differ significantly between \APRs and \HPRs. This suggests that once reviewers engage with \APRs, the additional effort needed to integrate them is similar to that required for human contributions. Overall, agentic coding provides a strong starting point that benefits from human oversight to ensure correctness, maintainability, and alignment with project conventions. In practice, developers can reduce review friction by keeping PRs small and encoding project-specific rules and architecture guidance for agents, while agent builders can integrate project-aware CI/CD checks and support automated maintenance tasks such as rebasing and conflict resolution.

\begin{acks}
We gratefully acknowledge the financial support of JSPS KAKENHI grants (JP24K02921, JP25K21359), as well as JST PRESTO grant (JPMJPR22P3), ASPIRE grant (JPMJAP2415), and AIP Accelerated Program (JPMJCR25U7). We also acknowledge the support of the Natural Sciences and Engineering Research Council of Canada (NSERC).
\end{acks}

\balance
\bibliographystyle{ACM-Reference-Format}
\bibliography{references}


\end{document}